\documentclass[journal]{IEEEtran}
\usepackage{ifpdf}
\usepackage{cite}
\ifCLASSINFOpdf
  \usepackage[pdftex]{graphicx}
\else
\fi
\usepackage{graphicx}
\usepackage{subfigure}
\usepackage{amsmath}
\usepackage{amssymb}
\usepackage{amsfonts}
\usepackage{amsmath,bm}
\usepackage{booktabs}
\usepackage{multirow}
\usepackage{algorithmic}
\usepackage{algorithm}
\usepackage{array}
\usepackage{xcolor}
\usepackage{booktabs}
\usepackage{tabularx}
\usepackage{array}
\usepackage{mathrsfs}
\usepackage{multirow, listings}
  \usepackage{verbatim}
\graphicspath{IEEEtran}

\usepackage{subfigure,cleveref}
\newcommand{\herm}{\mathsf{H}}

\begin{document}
% \bstctlcite{IEEEexample:BSTcontrol}

\title{SSB-Based Sensing-Assisted Robust Beamforming for High-Mobility UAV Communications in LAWN}
%\title{Robust Beamforming for SSB-Based Sensing-Assisted High-Mobility UAV Communications in ISAC Systems}

\author{Aimin~Tang,~\IEEEmembership{Senior~Member,~IEEE,}~Shuhan~Wang,~\IEEEmembership{Graduate~Student~Member,~IEEE,}~ Chongrui~Wang,~\IEEEmembership{Graduate~Student~Member,~IEEE,}~and ~Yin~Xu,~\IEEEmembership{Senior~Member,~IEEE}
\thanks{This work was supported in part by the Science and Technology Commission of Shanghai Municipality (Grant No. 25DP1500200). \\
Copyright (c) 2015 IEEE. Personal use of this material is permitted. However, permission to use this material for any other purposes must be obtained from the IEEE by sending a request to pubs-permissions@ieee.org.\\
The authors are all with Shanghai Jiao Tong University, Shanghai, 200240, China.  Corresponding author: Aimin Tang (email: tangaiming@sjtu.edu.cn).
}
}

\maketitle

\begin{abstract}
High-mobility uncrewed aerial vehicle (UAV) communications in low-altitude wireless networks (LAWN) demand reliable beamforming, while conventional feedback-based schemes suffer from excessive overhead and severe misalignment under rapid trajectory variations. To address this challenge, this paper proposes an SSB-based sensing-assisted predictive robust beamforming framework that replaces explicit CSI feedback with sensing-driven state estimation and uncertainty-aware optimization. Leveraging the periodic `always-on' synchronization signal block (SSB), a hierarchical sensing algorithm tailored for hybrid digital–analog uniform planar arrays is developed, combining 2D range-velocity profiling and augmented beamspace multiple signal classification (MUSIC). By integrating a locally-focused analog receive beamformer, the proposed sensing design can ensure energy accumulates across different radio-frequency (RF) chains while resolving angular ambiguity. An extended Kalman filter (EKF) is further employed to track UAV states between sparse synchronization-signal (SS) bursts, and a covariance correction is introduced to characterize maneuver-induced prediction uncertainties. Based on the derived statistical distributions of range and angular parameters, the communication channel is modeled through predictive correlation matrices rather than instantaneous CSI, leading to a multi-user robust beamforming formulation that maximizes average network sum-rate under uncertainty. The resulting nonconvex problem is efficiently solved via successive convex approximation and alternating minimization. Simulation results demonstrate that the proposed framework significantly enhances spectral efficiency and link stability compared with feedback-based beamforming and non-robust beamforming design, particularly in high-mobility and large-SSB-interval scenarios.
\end{abstract}

% \fi
\IEEEpeerreviewmaketitle

\section{Introduction}
Uncrewed aerial vehicles (UAVs) have emerged as a transformative technology, enabling a wide range of applications \cite{uav1,uav2}. To support reliable connections and effective management of UAVs, low-altitude wireless networks (LAWN) with both communication and sensing capabilities are indispensable \cite{yuan2025ground}. The emergence of integrated sensing and communications (ISAC) can offer a promising paradigm for UAV communications and sensing. By sharing hardware and spectrum, ISAC enables base stations (BSs) to implement sensing functionalities with minimal cost \cite{rcConvergence,survey,survey2,tang2024interference}. This integration allows for potential mutual enhancement: communication signals serve as the source for target perception, while sensing provides the geometric awareness to track fast-moving UAVs for robust communications. Besides the ISAC-based BS, UAVs in LAWN can also be enabled with ISAC function to provide both communication and sensing functions, where the energy efficiency, peak-to-average power ratio (PAPR), and trajectory optimization are crucial issues for UAV-enabled ISAC networks \cite{He2025Energy,Gu2025RIS}.

%energy efficiency is a critical concern for UAV-enabled ISAC due to limited onboard power. Beamforming and UAV trajectory are jointly optimized in \cite{He2025Energy} to maximize worst-case energy efficiency, achieving a favorable balance between sum rate and energy consumption under imperfect CSI.}

\begin{comment}
While UAVs offer remarkable flexibility and capabilities, their effective integration into the networks introduces new technical challenges. To ensure reliable communication and coordination, especially in dynamic environments, ground base stations (BSs) must establish and maintain robust links with UAVs \cite{uavConnectivity1}. This is particularly demanding in aerial scenarios, where long-range transmission and severe path loss significantly degrade signal quality. As a result, accurate and robust beamforming becomes a critical requirement for sustaining high-quality UAV-ground communication links in such high-mobility, long-distance conditions \cite{uavConnectivity2}.
\end{comment}
%To realize the potential of ISAC, a common approach adopted is to extract sensing information from the communication framework.

ISAC has been envisioned as a key technique for 6G networks, and 
the evolution of 6G networks adopts the communication framework to achieve sensing function \cite{22837}. Some experimental studies have been carried out to validate different sensing capabilities for 6G systems \cite{tang20266g,wang2026enabling}.    
While some literature directly utilizes communication data for sensing, e.g., \cite{data4sensing}, a more straightforward regime involves leveraging standardized reference signals (RS), such as demodulation reference signal (DMRS), positioning reference signal (PRS), and synchronization signal block (SSB), getting rid of the randomness of data. By utilizing these pre-existing RS structures, the system can estimate target parameters while preserving the legacy communication architecture. 
So far, extensive studies have been carried out on RS-based sensing. For example, some studies explore DMRS and PRS for sensing \cite{zhao2023reference,dmrs4sensing,prs4sensing1,prs4sensing2}. 
In addition to user-based DMRS and PRS, the public `always-on' SSB is one of the most promising candidates for sensing due to its periodic configuration and multi-beam-sweeping nature \cite{Li2025an,abrat2023ssb,ssbSIB1,Awad2025ssb}. For example, the studies in \cite{Li2025an,abrat2023ssb,ssbSIB1} explore the SSB for target detection under bistatic sensing mode. However, the SSB repetition interval (typically 20~ms or 40~ms) is too large to satisfy the Nyquist criterion for high-speed targets, causing high Doppler ambiguity. Therefore, SSB-based sensing via multiple SSB transmissions faces velocity ambiguity due to its large repetition interval. In \cite{ssbSIB1}, this problem is resolved by using additional downlink control information (DCI) and system information block 1 (SIB1). In \cite{Awad2025ssb}, monostatic sensing {(where the transmitter and receiver are co-located at the same BS)} with SSB is considered for multi-target detection under a digital beamforming architecture. 
These studies have demonstrated the feasibility of applying RS to detect UAVs in LAWN. However, effectively leveraging sensing results to assist robust UAV communications remains an open and interesting topic.

%{\color{blue}RIS-assisted waveform design has also emerged as a promising direction for UAV-ISAC. In \cite{Gu2025RIS}, joint waveform and RIS phase-shift optimization under PAPR constraints is investigated, where RIS is shown to be effective in mitigating multiuser interference while preserving sensing performance.}

Due to the high mobility of UAVs, maintaining stable links between ground BSs and UAVs becomes crucial \cite{uavConnectivity1,uavConnectivity2}. In multiple-input multiple-output (MIMO) systems, a robust beamforming design is essential to ensure reliable connectivity under such dynamic conditions. In this context, RS-based sensing serves as a vital source of prior information, enabling predictive beam alignment and tracking to reduce the overhead of conventional feedback-based schemes \cite{ssb1,ssb2}. % while maintaining robust connectivity between BSs and fast-moving UAVs. %The concept of sensing-assisted beam management reaches beyond UAV scenarios. Blockage detection and mobility detection are two critical functionalities that directly support and enhance the beam tracking procedure. In \cite{ssb1,ssb2}, SSBs are used to detect potential blockages, and other RSs are used to assess mobility and positional changes, both of which significantly improve beam alignment in mobile scenarios. In addition, the mobility state can be monitored to allocate dedicated beam training signals, providing more frequent updates for high-mobility users \cite{mobility}.
Intuitively, wireless sensing can provide angular domain information, which guides the design of the communication beam direction. For example, in \cite{colocatedRadar}, a radar colocated at the BS is used to estimate the user's angle of arrival (AoA), and the beamformer is designed to be the dominant eigenvector of the spatial covariance matrix. When the user's movement direction is known to the BS, which is often the case for vehicles on straight roads, angle prediction can be effectively carried out using extended Kalman filters (EKFs) \cite{radarVehicle}, Bayesian tracking methods \cite{radarVehicle2}, or orthogonal matching pursuit (OMP) algorithms \cite{radarVehicle3}, which helps maintain accurate beam alignment. 
Moreover, if the vehicle is modeled as an extended target, an adjustment of beamwidth is desired \cite{extendedTarget}. 
However, all the aforementioned studies assume that the road/motion direction is given. In contrast, when the user's motion direction is unknown to the BS, such as in the case of UAVs operating in three-dimensional and unconstrained spaces, other approaches are adopted. One common technique is to assume constant velocity motion, allowing EKF to be used for angle prediction over short time windows. EKF can be used to predict angles of UAVs, and the resource allocation can be optimized using reinforcement learning \cite{radarUav}. %Additionally, the multipath gain can be taken into account to find the optimal beam angle by using EKF with sensing information \cite{multipath}. 
However, for fast-moving UAVs, the information from sensing estimates and predictions can be erroneous or out-of-date. For example, if SSB is utilized to assist beamforming, the sensing information can be outdated to support the high-mobility UAV due to the large repetition interval of SSB. Furthermore, existing studies do not account for predictive beamforming design in multi-user settings, where co-channel interference, i.e., multi-user interference (MUI), significantly increases the problem complexity. Consequently, developing an effective and robust framework to handle such information and enable predictive multi-user beamforming remains an open challenge. %Moreover, these studies do not consider the predictive beamforming design for multiple users, where the multi-user interference (MUI) can make the problem more complicated. As a result, finding an effective and robust way of handling this information and developing predictive beamforming for multiple users remains an open problem.

%While many existing works on sensing-assisted beamforming focus on utilizing sensing information for better prediction of beam direction, they often overlook the co-channel interference among simultaneously served users. 

An effective beamforming design that reduces co-channel interference requires accurate channel state information (CSI) at the transmitter, which is often imperfect due to estimation errors, feedback delays, or quantization. 
To characterize the inevitable inaccuracies in imperfect CSI feedback, some existing studies on robust beamforming or precoding incorporate an error vector into the nominal channel representation \cite{robust1, robust2, robust3, robust4, slnr1, slnr2, robustTxRx}. This error vector is usually assumed to be either norm-bounded, spherically bounded, or drawn from a specific statistical distribution. Based on these assumptions, different types of formulation of robust optimization problems can be adopted.
In systems where power efficiency and interference suppression are critical, one can seek to minimize the total transmit power while guaranteeing that each user satisfies certain signal-to-interference-plus-noise ratio (SINR) or quality-of-service (QoS) constraints \cite{robust1, robust2, robust3, robust4}. 
Another approach utilizes the signal-to-leakage-plus-noise ratio (SLNR) as the performance metric, allowing the beamformer design problem to be decoupled across users \cite{slnr1,slnr2}. 
A third family of robust beamforming techniques is based on mean squared error (MSE), aiming to jointly optimize both downlink precoders and uplink combiners under imperfect CSI \cite{robustTxRx}. 
These approaches reflect different priorities and trade-offs in robust beamforming design under CSI uncertainty. However, they all need basic CSI feedback, and the assumptions of the error vector involved in the optimization problems can be further characterized if only sensing information is available. %, offering potential improvement in the overall performance.

In this paper, the sensing-assisted multi-user robust beamforming is considered. Particularly, the public, periodic SSB in 5G systems is utilized for sensing, and the obtained sensing information is further explored to assist predictive beamforming/beam-tracking. Specifically, a two-phase beamforming framework is established to ensure communication quality-of-service (QoS) in high-mobility scenarios. In the first phase, an SSB-based sensing algorithm tailored for hybrid digital-analog antenna structures is developed, enabling the periodic collection of positional information. In the second phase, predictive beamforming/beam-tracking is conducted based on an EKF to track UAV states. Compared to the requirements of beam-tracking frequency, the sensing measurements from SSB are very sparse, which cannot be directly applied to support accurate beamforming over a long duration between two SSBs. Therefore, the prediction uncertainty of the EKF is further characterized in terms of estimated range and angles. This statistical uncertainty is then leveraged to formulate a multi-user robust beamforming optimization problem, which maximizes the network's average sum-rate while ensuring resilience against mobility-induced errors. Simulations are finally conducted to demonstrate the effectiveness of the proposed sensing algorithm and beamforming performance.

\begin{figure*}[t]
    \centering
    \includegraphics[width = 0.75\textwidth]{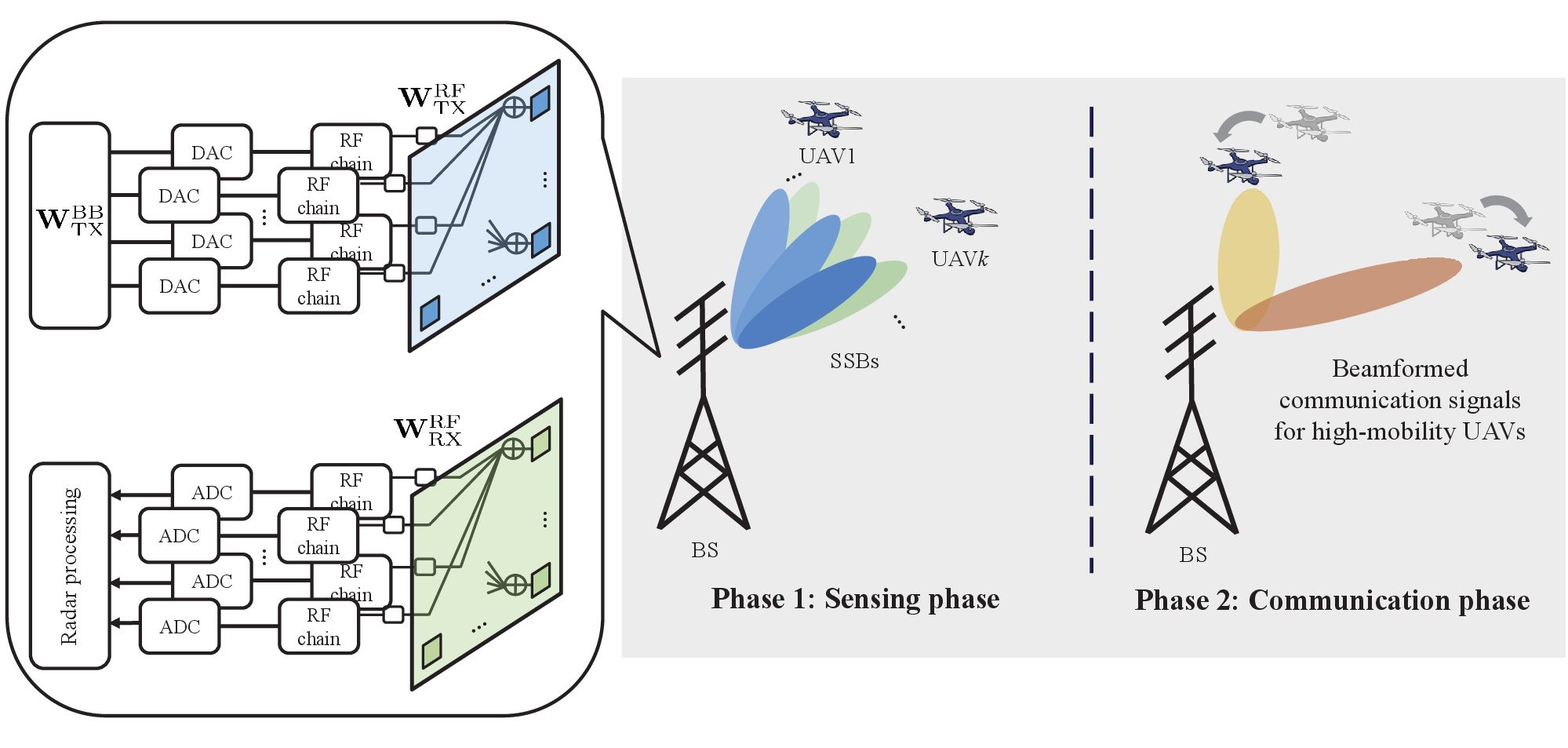}
    \vspace{-1em}
    \caption{System architecture and an illustration of SSB-based sensing-assisted beamforming for high-mobility UAV communications.}
    \label{fig:setup}
    \vspace{-1em}
\end{figure*}

The contributions of this paper are summarized as follows:
\begin{itemize}
    \item An SSB-based sensing-assisted robust beamforming framework is established for multi-user UAV scenarios. Two iterative phases, namely, the sensing phase and the communication phase, are structured to replace the conventional feedback procedure for beamforming and beam realignment. The SSB is utilized for sensing, while an EKF is adopted to predict the movements of UAVs to further assist beamforming/beam-tracking. Due to the large repetition interval of SSB, the sensing information can be outdated for fast-moving UAVs. Thus, the prediction with uncertainty of the EKF is incorporated into the framework to support robust beamforming. The multi-UAV robust beamforming can finally be achieved by utilizing the prior information obtained from the SSB-based sensing phase with uncertainty.
    \item An SSB-based sensing algorithm is developed for ISAC systems with hybrid digital-analog uniform planar array (UPA) structures. In order to enhance the sensing capability provided by SSB, a hierarchical sensing scheme is developed. The 2D discrete Fourier transform (DFT) is first adopted on the delay-Doppler (DD) domain for effective target detection and range-velocity estimation. Based on the range-velocity profiling, the angle information is further obtained by leveraging 2D augmented beamspace MUSIC. Particularly, a locally-focused receive analog beamformer is designed for SSB-based sensing, which can resolve the angular ambiguity problem inherent in hybrid architectures while accumulating the energy across different radio-frequency (RF) chains. A tracking scheme based on EKF is further established to frequently obtain user-related parameters between SSB repetitions. The characterization of potential tracking errors of the EKF is derived through the corrected state covariance, which can serve as statistical information for robust beamforming.
    \item In contrast to prior works that assume known motion directions and focus solely on single-user predictive beamforming, this paper develops a predictive multi-user robust beamforming design that integrates both the uncertainty statistics from the EKF and the MUI into the SINR constraints. By modeling the channel as a function of the statistical error distribution rather than a nominal point estimate, the beamforming design effectively accounts for angular and positional uncertainties. %The MUI is also addressed in this model. 
    The resulting non-convex optimization problem is solved using successive convex approximation (SCA) and alternating minimization. The performance of our proposed scheme can significantly outperform the feedback-based solution for high-mobility UAVs, due to lower overhead and more robust connectivity.
\end{itemize}

The rest of this paper is organized as follows. The system and signal models are presented in Section \ref{sec:model}. The sensing-assisted beamforming framework is provided in Section \ref{sec:framework}. The SSB-based sensing signal processing is elaborated in Section \ref{sec:sensing}, followed by the EKF tracking procedure in Section \ref{sec:tracking}, and multi-user beamforming in Section \ref{sec:beamforming}. The simulations are carried out in Section \ref{sec:sim}, and this paper is concluded in Section \ref{sec:conclusion}. 

\textbf{Notations:} $x$, $\mathbf{x}$, $\mathbf{X}$ represent a scalar, a vector, and a matrix. $(\mathbf{X})_{i,j}$ is the $i$-th row and $j$-th column of $\mathbf{X}$. $\mathbf{x}^\mathsf{T}$ is the transpose of $\mathbf{x}$ and $\mathbf{X}^\herm$ is the Hermitian transpose of $\mathbf{X}$. Operators "$\otimes$" and "$\odot$" represent the Kronecker product and element-wise multiplication, respectively. Function $\mathrm{arctan2}(y, x)$ denotes the four-quadrant inverse tangent function. Operator $\mathbf{X}^\dagger$ is the pseudo-inverse of $\mathbf{X}$.
The main symbols used throughout this paper are summarized in Table~\ref{tab:key_notations}.

\begin{table}[t]
\centering
\caption{Key Notations}
\label{tab:key_notations}
\scriptsize
\renewcommand{\arraystretch}{1.08}
\setlength{\tabcolsep}{1.5pt}
\begin{tabularx}{\columnwidth}{
@{}
>{\centering\arraybackslash}p{0.38\columnwidth}
>{\raggedright\arraybackslash}X
@{}}
\toprule
\textbf{Symbol} & \textbf{Definition} \\
\midrule
$K$
& Number of UAVs \\
$N_{\mathrm{TX}},\,N_{\mathrm{RX}}$
& Numbers of BS TX/RX antennas \\
$N_{\mathrm{TX}}^{\mathrm{RF}},\,N_{\mathrm{RX}}^{\mathrm{RF}}$
& Numbers of TX/RX RF chains \\
$\mathbf{W}_{\mathrm{TX,com}}^{\mathrm{RF}},\,
 \mathbf{W}_{\mathrm{TX,com}}^{\mathrm{BB}}$
& Communication analog/digital beamformers \\
$\mathbf{w}_{\mathrm{TX,com},k}$
& Communication beamformer for UAV $k$ \\
$\mathbf{h}_k$
& Downlink channel vector of UAV $k$ \\
$\mathbf{S}_{\mathrm{SSB}}$
& Frequency-domain SSB signal matrix \\
$\mathbf{W}_{\mathrm{TX,SSB}}^{\mathrm{RF}},\,
 \mathbf{w}_{\mathrm{TX,SSB}}^{\mathrm{BB}}$
& SSB analog/digital transmit beamformers \\
$\mathbf{W}_{\mathrm{RX,rad}}^{\mathrm{RF}}$
& Sensing analog receive beamformer \\
$r_k,\,v_k$
& Range/radial velocity of UAV $k$ \\
$\phi_k,\,\theta_k$
& Elevation/azimuth angles of UAV $k$ \\
$T,\,T_{\mathrm{KF}}$
& SS-burst/EKF prediction intervals \\
$\mathbf{m}_k^{(i)},\,\mathbf{s}_k^{(i)}$
& EKF measurement/state vectors of UAV $k$ \\
$\mathbf{P}_k^{(i|i-1)},\,\mathbf{P}_k^{(i|i)}$
& Predicted/updated EKF covariances \\
$\widetilde{\mathbf{P}}_k^{(i)}$
& Corrected state covariance matrix \\
$\sigma_{r,k}^{2},\,\sigma_{\phi,k}^{2},\,\sigma_{\theta,k}^{2}$
& Range/elevation/azimuth error variances \\
$\mathbf{R}_k$
& Channel correlation matrix of UAV $k$ \\
$C_{r,k},\,\mathbf{R}_{\phi,\theta,k}$
& Path-loss factor/angular correlation matrix \\
$\sigma_{\mathrm{eff},k}^{2}$
& Effective noise variance of UAV $k$ \\
$\gamma_k$
& SINR requirement of UAV $k$ \\
$P_{\max}$
& Maximum transmit power \\
\bottomrule
\end{tabularx}
\vspace{-1em}
\end{table}

\section{System and Signal Models}\label{sec:model}
\subsection{System Model}
The system architecture is illustrated in Fig. \ref{fig:setup}, where a BS with ISAC function senses the aerial environment using SSB signals and communicates with $K$ fast-moving single-antenna UAVs in suburban or relatively open LAWN environments. The BS is equipped with an $N_{\rm{TX}}$-antenna transmit UPA and an $N_{\rm{RX}}$-antenna receive UPA. %In order to have greater flexibility in beam pattern design and higher array gain,
The fully-connected hybrid architecture is adopted for UPAs, with the transmit (receive) UPA connected to $N_{\rm{TX}}^{\rm{RF}}$ ($N_{\rm{RX}}^{\rm{RF}}$) RF chains. The transmit UPA consists of $N_{\rm{TX,y}}$ antenna elements horizontally, and $N_{\rm{TX,z}}$ elements vertically, with $N_{\rm{TX,y}}N_{\rm{TX,z}} = N_{\rm{TX}}$. There are $N_{\rm{TX,y}}^{\rm{RF}}$ horizontal RF chains and $N_{\rm{TX,z}}^{\rm{RF}}$ vertical RF chains at transmit UPA, with $N_{\rm{TX,y}}^{\rm{RF}}N_{\rm{TX,z}}^{\rm{RF}} = N_{\rm{TX}}^{\rm{RF}}$. Similarly, the receive UPA is composed of $N_{\rm{RX,y}}$ and $N_{\rm{RX,z}}$ antenna elements horizontally and vertically, connecting to $N_{\rm{RX,y}}^{\rm{RF}}$ and $N_{\rm{RX,z}}^{\rm{RF}}$ horizontal and vertical RF chains.

\subsection{Signal Model for Communications}
In this paper, the downlink transmission from the BS to multiple UAVs is considered. Assume that the communication data transmitted to the $k$-th user at the $n$-th subcarrier and $m$-th OFDM symbol in the frequency domain is $s_k[n,m]$. For notational simplicity, the subcarrier index and symbol index are dropped. Denote $\mathbf{s} = [s_1, s_2, \cdots, s_K]^{\mathsf{T}}$ as the transmitted signal vector. %, whose elements are assumed to be independent and identically distributed (i.i.d.) with unity variance, i.e. $\mathbb{E}[\mathbf{s}\mathbf{s}^{\mathsf{H}}] = \mathbf{I}$. 
The transmit analog beamformer is denoted as $\mathbf{W}_{\rm{TX,com}}^{\rm{RF}}\in \mathbb{C}^{N_{\rm{TX}}\times N_{\rm{TX}}^{\rm{RF}}}$, and the transmit digital beamformer is denoted as $\mathbf{W}_{\rm{TX,com}}^{\rm{BB}}\in \mathbb{C}^{N_{\rm{TX}}^{\rm{RF}} \times K}$, which can be written as $\mathbf{W}_{\rm{TX,com}}^{\rm{BB}} = [\mathbf{w}_{\rm{TX,com},1}^{\rm{BB}},\mathbf{w}_{\rm{TX,com},2}^{\rm{BB}},\cdots, \mathbf{w}_{\mathrm{TX,com}, K}^{\rm{BB}}]$. The analog beamformer is implemented by phase shifters, i.e., 
\begin{equation}
    |(\mathbf{W}_{\rm{TX,com}}^{\rm{RF}})_{i,j}| \!=\! \frac{1}{\sqrt{N_{\rm{TX}}}},\, \forall i \in [1,N_{\rm{TX}}], j\in [1,N_{\rm{TX}}^{\rm{RF}}].
    \label{eq:WRF}
\end{equation}
The transmit communication beamforming vector for the $k$-th user is denoted as $\mathbf{w}_{\mathrm{TX,com},k} = \mathbf{W}_{\rm{TX,com}}^{\rm{RF}} \mathbf{w}_{\mathrm{TX,com},k}^{\rm{BB}} \in \mathbb{C}^{N_{\rm{TX}}\times 1}$. Combining all beamforming vectors forms the beamforming matrix $\mathbf{W}_{\mathrm{TX,com}} = \mathbf{W}_{\rm{TX,com}}^{\rm{RF}} \mathbf{W}_{\rm{TX,com}}^{\rm{BB}}= [\mathbf{w}_{\mathrm{TX,com},1}, \mathbf{w}_{\mathrm{TX,com},2}, \cdots, \mathbf{w}_{\mathrm{TX,com},K}]$. Thus, the beamformed transmitted signal can be represented as 
\begin{equation}
    \mathbf{x}_{\mathrm{com}} = \sum_{k=1}^K \mathbf{w}_{\mathrm{TX,com},k} s_k = \mathbf{W}_{\mathrm{TX,com}}\mathbf{s}.
\end{equation}

The channel for the $k$-th user can be modeled as 
\begin{equation}
    \mathbf{h}_k = \sum_{l=1}^{L_k} \alpha_{k,l} \mathbf{a}_{k,l}(\phi_{k,l},\theta_{k,l})\in \mathbb{C}^{N_{\rm{TX}}\times 1}, 
\end{equation}
where $L_k$ is the total number of paths, $\alpha_{k,l}$, $\phi_{k,l}$, and $\theta_{k,l}$ are the complex path gain, elevation angle, and azimuth angle for the $l$-th path of the $k$-th user, respectively. The path gain and angles are modeled as random variables that follow certain distributions within a short coherent-time period. $\mathbf{a}(\phi,\theta) = \mathbf{a}_{\rm{z}}(\phi) \otimes \mathbf{a}_{\rm{y}}(\phi,\theta)$ is the 2D steering vector, where $\mathbf{a}_{\rm{z}}(\phi) = [1, e^{j 2 \pi d_{\rm{z}} \text{sin}(\phi)/\lambda}, \cdots e^{j 2 \pi (N_{\rm{z}}-1) d_{\rm{z}} \text{sin}(\phi)/
\lambda}]^\mathsf{T}$ and $\mathbf{a}_{\rm{y}}(\phi,\theta) = [1, e^{j 2 \pi d_{\rm{y}} \text{sin}(\theta)\text{cos}(\phi)/\lambda}, \cdots e^{j 2 \pi (N_{\rm{y}}-1) d_{\rm{y}} \text{sin}(\theta)\text{cos}(\phi)/
\lambda}]^\mathsf{T}$. $d_{\mathrm{y}}$ and $d_{\mathrm{z}}$ are antenna spacing at $y$-axis and $z$-axis respectively, and $\lambda$ denotes the wavelength. Generally, the element spacings are set to half the wavelength.
Since this work focuses on low-altitude UAV communications in suburban or relatively open LAWN environments, where the BS-to-UAV links are expected to be LoS-dominant, the LoS component is considered the dominant propagation path. Therefore, in the following analytical beamforming design, we set $L_k = 1,\forall k$. %This assumption is mainly applicable to suburban environments or areas with relatively low-rise buildings. 
In dense urban environments with high-rise buildings and strong reflections, NLoS and multipath components may become significant, which is beyond the scope of this work and requires further investigation. The velocity/Doppler effect is not explicitly incorporated into the channel model. This is attributed to the block-by-block decoding strategy operating at a sub-millisecond granularity, within which phase variations are adequately compensated for by leveraging DMRS-aided channel estimation. Furthermore, over radar processing intervals spanning several milliseconds, the resulting displacement of even highly maneuverable UAVs is minimal, rendering the associated path-gain fluctuations insignificant.

%The impact of velocity/Doppler effect for each user is not explicitly modeled in our communication channel model. This is because the communication decoding is conducted in a block-by-block way in a very small time scale, such as sub-microsecond. With the help of DMRS for channel estimation, the phase change within a communication block can be well tracked. %Our previous study has shown how sensing information can be utilized to accurately assist the communication channel estimation under high mobility \cite{CE}. Furthermore, in large-scale radar processing intervals, such as a few milliseconds, the position change of high-mobility UAV is still very small, so that the impact of high mobility on path gain can also be ignored. 

%Given the channel model, the received signal at the $k$-th user is given by
The received signal at the $k$-th user is given by
\begin{equation}
    \begin{aligned}
        y_{\mathrm{com},  k} 
       % &= \mathbf{h}_k^{\mathsf{H}} \mathbf{x}_{\mathrm{com}} + z_k \\&
        = \mathbf{h}_k^{\mathsf{H}} \mathbf{w}_{\mathrm{TX,com},k} s_k + \mathbf{h}_k^{\mathsf{H}} \!\!\sum_{i = 1,i\neq k}^K \!\!\!\mathbf{w}_{\mathrm{TX,com},i} s_i + z_k,
    \end{aligned}
\end{equation}
%where $z_k$ is the noise element for the $k$-th user and $z_k \sim \mathcal{CN}(0,\sigma_{\mathrm{z},k}^2)$. 
where $z_k \sim \mathcal{CN}(0,\sigma_{\mathrm{z},k}^2)$ is the noise element for the $k$-th user.

\subsection{Signal Model for Sensing}
\begin{figure}[t]
    \centering
    \includegraphics[width=0.4\textwidth]{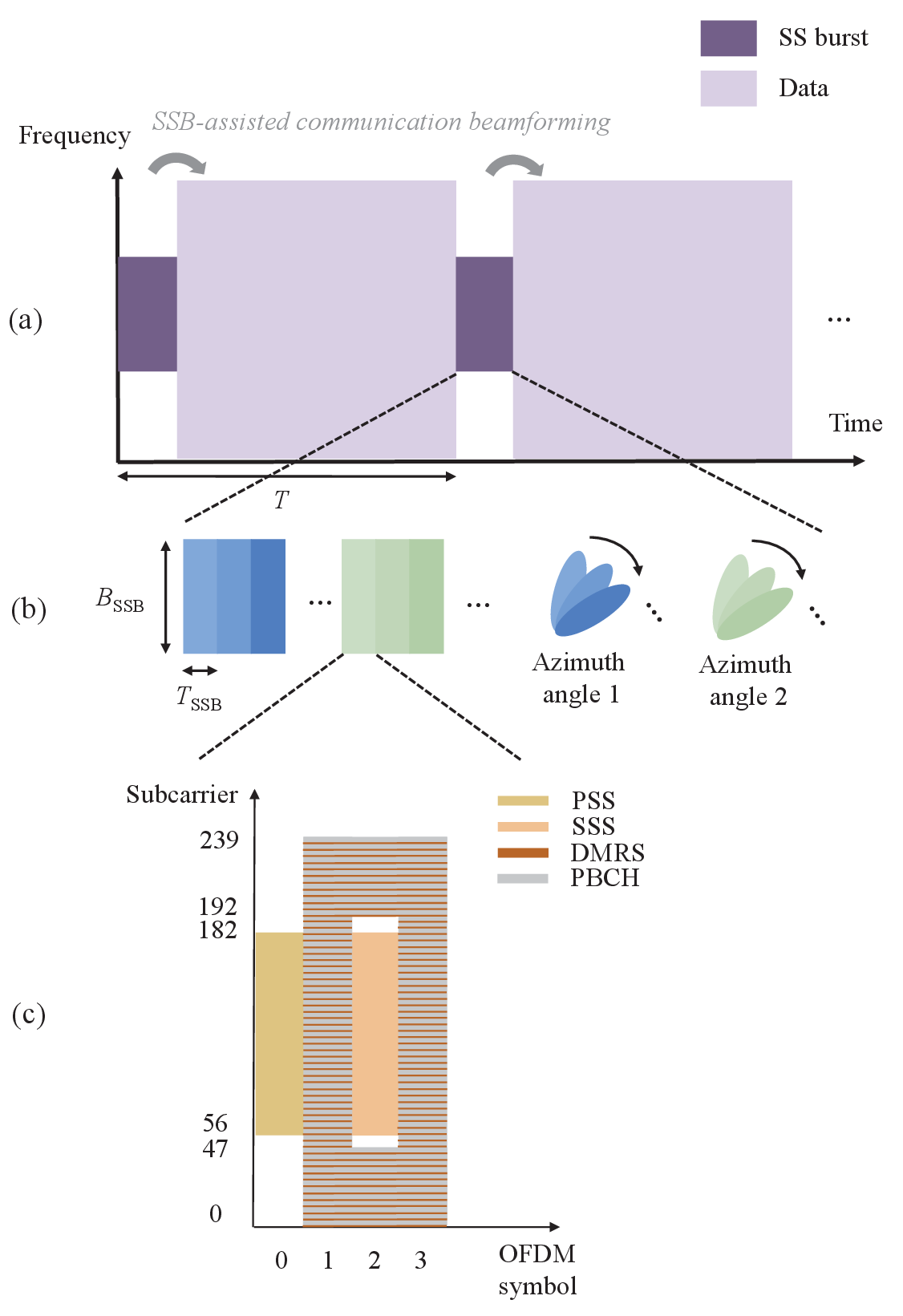}
    \caption{An illustration of the SSB-assisted communication beamforming frame structure. (a) SSB sensing-assisted communication beamforming frame structure; (b) SS burst structure; (c) SSB frame structure.}
    \vspace{-0em}
    \label{fig:ssb_frame_structure}
    \vspace{-1em}
\end{figure}

In 5G networks, the SSB can be a leading candidate for sensing applications due to its `always-on' periodic transmission and beam-sweeping capability. The BS transmits a sequence of SSBs, with each steered toward a distinct spatial direction. Together, these SSBs form a synchronization signal (SS) burst for sweeping the entire coverage area. The physical structure of the SSB follows the 5G NR specifications \cite{38213}. Specifically, the BS sequentially transmits multiple SSBs, each steered to a distinct spatial direction to cover the whole angular region, as shown in Fig. \ref{fig:ssb_frame_structure}(b). The repetition period of SS burst can be configured from a set of [5 10 20 40 80 160] ms. Shorter periods (e.g., 5 ms) allow for faster cell acquisition by users, while longer periods (e.g., 160 ms) improve network efficiency. An SSB occupies 4 consecutive OFDM symbols in time and 240 subcarriers in frequency (equivalently 20 physical resource blocks), as shown in Fig. \ref{fig:ssb_frame_structure}(c). The SSB payload comprises the primary synchronization signal (PSS), the secondary synchronization signal (SSS), and the physical broadcast channel (PBCH), together with the associated PBCH DMRS. The SSB's occupation of the time-frequency resource elements is not continuous, which leaves the non-occupied resource elements zero.

In this paper, we consider that the BS uses the echo of SSBs for sensing, i.e., the BS performs monostatic sensing by utilizing SSBs. $N$ is the number of subcarriers occupied by SSB, and $M$ is the number of OFDM symbols within one SSB. For a given SSB, the transmit signal at the $n$-th subcarrier and the $m$-th OFDM symbol is beamformed as
\begin{equation}
    \mathbf{x}_{\mathrm{SSB},n,m} = \mathbf{W}_{\mathrm{TX,SSB}}^{\rm{RF}} \mathbf{w}_{\mathrm{TX,SSB}}^{\rm{BB}} (\mathbf{S}_{\rm{SSB}})_{n,m},
\end{equation}
where $\mathbf{W}_{\mathrm{TX,SSB}}^{\rm{RF}} \in \mathbb{C}^{N_{\rm{TX}}\times N_{\mathrm{TX}}^{\rm{RF}}}$ is the analog transmit beamformer, $\mathbf{w}_{\mathrm{TX,SSB}}^{\rm{RF}} \in \mathbb{C}^{ N_{\mathrm{TX}}^{\rm{RF}} \times 1}$ is the digital transmit beamformer, and $\mathbf{S}_{\rm{SSB}} \in \mathbb{C}^{N\times M}$ is the frequency-domain signal of the transmitted SSB. Note that the SSB transmit beamformer $\mathbf{W}_{\mathrm{TX,SSB}}^{\rm{RF}} \mathbf{w}_{\mathrm{TX,SSB}}^{\rm{BB}}$ is a predefined vector that points at the desired SSB direction for cell acquisition by
users.

%Assume that the $l$-th sensing target detected during the SSB, i.e., the $l$-th UAV, is characterized by $r_l$, $v_l$, $\phi_l$, and $\theta_l$, denoting its distance, velocity, elevation angle, and azimuth angle with respect to the BS, respectively. 
The sensing channel perceived by the BS is modeled as
\begin{equation}
    \mathbf{H}_{\mathrm{rad},n,m} \!\!= \!\!\sum_{l=1}^{L}\alpha_l e^{-j 2 \pi n \Delta_\mathrm{f} \tau_l} e^{j 2 \pi m T_{\mathrm{o}}f_{\mathrm{d},l}} \mathbf{a}(\phi_l,\theta_l) \mathbf{a}(\phi_l,\theta_l)^\mathsf{H}\!\!,
\end{equation}
where $L$ is the number of UAVs detected within the area illuminated by the current SSB. $\alpha_l = \sqrt{\frac{c_0^2 \sigma_l}{(4\pi)^3 r_l^4 f_\mathrm{c}^2}}$, $\tau_l = \frac{2 r_l}{c_0}$, $f_{\mathrm{d},l} = \frac{2 v_l f_\mathrm{c}}{c_0}$, $\phi_l$, and $\theta_l$ are the two-way attenuation, delay, Doppler shift, elevation angle, and azimuth angle for the $l$-th target, respectively. $r_l$ and $v_l$ are the range and radial velocity of the $l$-th target. $\sigma_l$ is the RCS of the $l$-th target, $\Delta_{\mathrm{f}}$ is the subcarrier spacing, $T_\mathrm{o}$ is the duration of the OFDM symbol including the cyclic prefix, $c_0$ is the speed of light, and $f_\mathrm{c}$ is the carrier frequency.

Since a hybrid beamforming architecture is considered at the BS receiver side, the sensing received signal at the BS after analog beamforming can thus be expressed as
\begin{equation}
\begin{aligned}
    \mathbf{y}_{\mathrm{rad},n,m} \! &= (\mathbf{W}_{\mathrm{RX,rad}}^{\rm{RF}})^{\mathsf{H}} \mathbf{H}_{\mathrm{rad},n,m} \mathbf{x}_{\mathrm{SSB},n,m} + \mathbf{z}'_{n,m}\\
    %& = \mathbf{W}_{\mathrm{RX,rad},n,m}^{\rm{RF}} \mathbf{H}_{\mathrm{rad},n,m}\mathbf{W}_{\mathrm{TX,SSB},n,m}^{\mathrm{RF}} \mathbf{w}_{\mathrm{TX,SSB},n,m}^{\rm{BB}} (\mathbf{S}_{\rm{SSB}})_{n,m} \!\!+ \!\mathbf{z}'_{n,m} \\
    & = \mathbf{h}_{\mathrm{rad},n,m} (\mathbf{S}_{\rm{SSB}})_{n,m} + \mathbf{z}'_{n,m},
\end{aligned}
\end{equation}
where $\mathbf{h}_{\mathrm{rad},n,m} = (\mathbf{W}_{\mathrm{RX,rad}}^{\rm{RF}})^{\mathsf{H}} \mathbf{H}_{\mathrm{rad},n,m}\mathbf{W}_{\mathrm{TX,SSB}}^{\rm{RF}} \mathbf{w}_{\mathrm{TX,SSB}}^{\rm{BB}}\in \mathbb{C}^{N_{\rm{RX}}^{\rm{RF}}\times 1}$ is the effective sensing channel, $\mathbf{W}_{\mathrm{RX,rad}}^{\rm{RF}} \in \mathbb{C}^{N_{\rm{RX}}\times N_{\rm{RX}}^{\rm{RF}}}$ is the receive analog beamformer, and $\mathbf{z}'_{n,m}$ is the white Gaussian noise vector. 
The digitally received signal $\mathbf{y}_{\mathrm{rad},n,m}$ is further used for radar sensing, while the analog beamformer $\mathbf{W}_{\mathrm{RX,rad}}^{\rm{RF}}$ should be carefully designed to enhance sensing performance.

\section{Framework Design for Sensing-Assisted UAV Beamforming} \label{sec:framework}
\begin{figure}[t]
    \centering
    \includegraphics[width=0.45\textwidth]{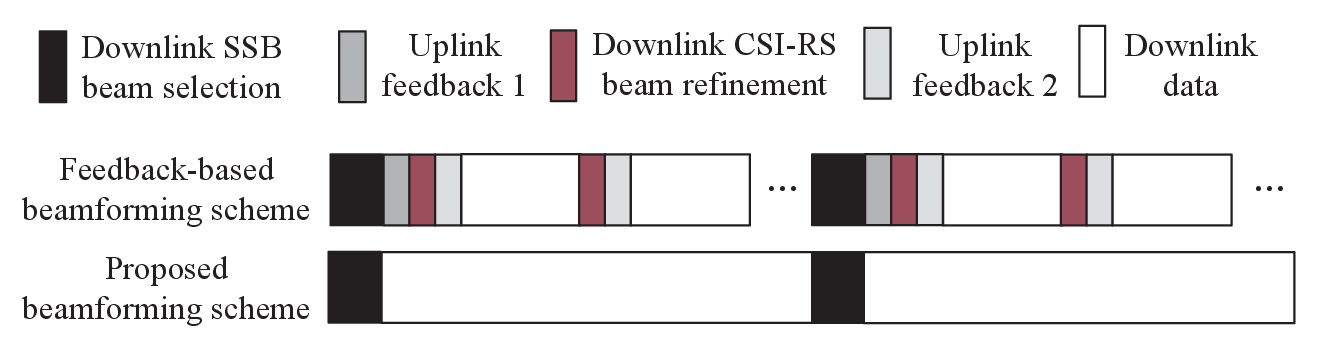}
    \caption{Comparison between the feedback-based beamforming scheme and the proposed sensing-assisted beamforming scheme.}
    \label{fig:schemeComp}
    \vspace{-1em}
\end{figure}

To support robust communication with high-mobility UAVs, beam training and tracking are critical for large MIMO arrays. However, the process of conventional beam training and tracking suffers from high overhead for fast-moving UAVs.  
%Due to the high mobility of UAVs and the high angular resolution required by large MIMO arrays, the signaling overhead inherent in the conventional beam training scheme becomes large. 
As illustrated in Fig. \ref{fig:schemeComp}, the conventional beam training scheme involves a ``coarse-to-fine" sweeping and feedback mechanism. The BS first broadcasts SSBs for wide-beam sweeping, after which each UAV reports its preferred beam index based on received signal strength. Next, for each user's selection of the wide beam, the BS conducts narrow-beam sweeping using channel state information reference signal (CSI-RS) for refinement. %This can introduce large overhead, while lacking consideration of inter-user interference for multi-user scenarios. 
In the beam tracking process, the BS periodically refines the beam for each user with the sweeping of CSI-RS. The overhead of beam refinement is scaled linearly with the number of users.  
%Moreover, even with the CSI-RS refinement, 
Furthermore, since the beamformer remains static between sweeping intervals, the high mobility of UAVs inevitably leads to beam misalignment, as the fixed beamformer lacks the robustness to accommodate rapid trajectory changes. In contrast, the sensing-assisted scheme can bypass the feedback procedure by directly extracting user-related parameters from sensing, significantly reducing the overhead while increasing the effective data transmission window through predictive beamforming.

In this paper, we develop an SSB-based sensing-assisted robust beamforming framework. Only the `always-on' SSB is required, and thus, the overhead for the refinement and beam tracking in the conventional procedure can be eliminated. 
%In order to realize sensing-assisted multi-user UAV beamforming, a systematic framework is established. 
The proposed framework at the BS is structured into two iterative phases as illustrated in Fig. \ref{fig:ssb_frame_structure}(a). 
The framework employs SSB for sensing and an EKF for predictive tracking of UAV kinematics, which aids beamforming and beam-tracking. To mitigate the performance degradation caused by the SSB's large repetition interval, which leads to stale sensing updates for fast-moving UAVs, the statistical uncertainty of the EKF's state estimates is incorporated into the beamformer optimization. Consequently, robust multi-UAV beamforming is achieved by utilizing the prior information from the SSB-based sensing phase, explicitly accounting for its uncertainty. %The details of the two phases are elaborated as follows.
%A core design focus of this framework lies in the design of two distinct beamformers, namely, the sensing receive analog beamformer $\mathbf{W}_{\mathrm{RX,rad}}^\mathrm{RF}$ and the multi-user communication transmit beamformer $\mathbf{W}_{\mathrm{TX,com}}$. 

In Phase 1 (sensing phase), the BS utilizes SSBs, the periodically transmitted downlink RS with a wide-beam sweeping for the whole-area coverage, not only for initial access, but also as active probing signals. While the sensing transmit beamforming is pre-determined by the SSB sweeping patterns, the design of the receive analog beamformer $\mathbf{W}_{\mathrm{RX,rad}}^\mathrm{RF}$ is required for the sensing algorithm. By performing monostatic sensing with the reflected SSB signals, the BS obtains estimates of the UAV parameters, including ranges, velocities, elevation angles, and azimuth angles. This provides the system with environmental priors and real-time awareness of the individual channel states without relying on user-side CSI reports. 

In Phase 2 (communication phase), to maintain robust communication services between sparse SS bursts with interval $T$, the sensing-derived information is integrated into an EKF-based tracking mechanism. The EKF operates at a finer granularity, providing continuous state predictions. 
The sparse measurements from the SS burst lead to prediction errors in the intervals between them. To achieve robust beamforming in the communication phase, we derive a characterization of this prediction uncertainty and incorporate it into the beamforming design.
%Due to the measurement from the SS burst being sparse, the prediction in between can have some errors without the measurement. Targeted at robust beamforming in the communication phase, the characterization of the prediction uncertainty is derived for robust beamforming. 
Specifically, the measurement obtained from the SSB-based sensing algorithm is fed to the EKF with a period of $T$, while the EKF states are updated to get a predicted state at $T_\mathrm{KF}$ to support communication beamforming between two SS bursts. %Thus, the prediction uncertainty should be considered. 
The communication hybrid beamformer $\mathbf{W}_{\mathrm{TX,com}}$ is dynamically optimized based on the prediction uncertainty derived from the EKF to ensure high-gain, interference-aware communication that is robust against the tracking latencies and maneuvers inherent in high-mobility aerial scenarios. 

\section{Sensing Signal Processing} \label{sec:sensing}
At the BS, the target parameters $(r_l, v_l, \phi_l, \theta_l)$ are estimated from the received frequency-domain SSB signals for every SS burst. However, the joint estimation of these four parameters poses significant challenges, including the high computational complexity of a simultaneous four-dimensional search and the angular ambiguity inherently caused by the hybrid antenna architecture. To address the angular ambiguity, existing literature employs subspace-based techniques such as beamspace MUSIC/ESPRIT \cite{beamspace1, beamspace2}, or cross-correlation approaches that require a fixed phase shift between adjacent subarrays \cite{coherentAccu, spatialFilter}. Nevertheless, the latter is strictly limited to the array-of-subarray structure and is inapplicable to the fully-connected hybrid setup assumed in this paper. 

To overcome the aforementioned challenges, a hierarchical sensing algorithm is developed, where range and velocity $(r_l, v_l)$ are first estimated using the periodogram-based method and azimuth and elevation $(\phi_l, \theta_l)$ are then estimated with 2D beamspace MUSIC with a carefully designed analog beamformer.

%The range and velocity $(r_l, v_l)$ are first estimated using the periodogram-based method. To bridge these results with the spatial domain, 2D beamspace MUSIC for azimuth and elevation $(\phi_l, \theta_l)$ estimation is performed for each detected range-velocity peak by gathering signal snapshots within a local window around the peak. To resolve the angular ambiguity problem inherent in hybrid architectures while accumulating the energy across different RF chains, a locally-focused receive analog beamformer is designed for SSB-based sensing.

\subsection{Range-Velocity Profiling}
The sensing channel is first extracted. The least-squares (LS) estimate of channel frequency response (CFR) at the $n$-th subcarrier and $m$-th OFDM symbol can be obtained by element-wise division as
\begin{equation}
\begin{aligned}
       \hat{\mathbf{h}}_{\mathrm{rad},n,m}=\left\{\begin{array}{ll}
                \frac{\mathbf{y}_{\mathrm{rad},n,m}}{(\mathbf{S}_{\mathrm{SSB}})_{n,m}}, & \text { if } (n,m) \in\Omega\\
                0, & \text { if } (n,m) \notin \Omega
        \end{array}, \right. \\
\end{aligned}
\end{equation}
where $\Omega$ is the resource element set that SSB occupies.

2D DFT is then implemented for periodogram-based range-velocity estimation. The inverse DFT (IDFT) is applied along the subcarrier to obtain the delay profile as
\begin{equation}
    \mathbf{d}_{i,m} = \sum_{n=1}^{N_{\mathrm{IDFT}}} q_{i,n}\hat{\mathbf{h}}_{\mathrm{rad},n,m}' = [\hat{\mathbf{h}}_{\mathrm{rad},1,m}', \cdots \hat{\mathbf{h}}_{\mathrm{rad},N_{\mathrm{IDFT}},m}']\mathbf{q}_i,
\end{equation}
where $i \in [1,N_{\rm{IDFT}}]$, $N_{\rm{IDFT}}$ is the number of IDFT points applied along subcarrier, $\mathbf{q}_i \in \mathbb{C}^{N_{\rm{IDFT}}\times 1}$ is the IDFT vector with each element given by $(\mathbf{q}_i)_n = q_{i,n} = e^{j2\pi ni/N_{\rm{IDFT}}}$, and $\hat{\mathbf{h}}_{\mathrm{rad},n,m}'$ is the zero-padded version of $\hat{\mathbf{h}}_{\mathrm{rad},n,m}$, where zeros are added to the subcarriers given $N_{\rm{IDFT}} \geq N$. 

Similarly, DFT can be applied along the OFDM symbols to further obtain the delay-Doppler profile as
\begin{equation}
    \mathbf{v}_{i,j} = \sum_{m = 1}^{M_{\mathrm{DFT}}}q'_{j,m}\mathbf{d}'_{i,m} = [\mathbf{d}'_{i,1},\cdots,\mathbf{d}'_{i,M_{\rm{DFT}}}]\mathbf{q}'_j,
\end{equation}
where $j \in [1,M_{\rm{DFT}}]$, $M_{\rm{DFT}}$ is the number of DFT points applied along symbol, $\mathbf{q}'_i \in \mathbb{C}^{M_{\rm{DFT}}\times 1}$ is the DFT vector with each element given by $(\mathbf{q}'_j)_m = q'_{j,m} = e^{-j2\pi mj/M_{\rm{DFT}}}$, and $\hat{\mathbf{d}}_{i,m}'$ is the zero-padded version of $\hat{\mathbf{d}}_{i,m}$, where zeros are added to the OFDM symbols given $M_{\rm{DFT}} \geq M$. 

The delay-Doppler (DD) profile $\mathbf{v}_{i,j} = [v_{1,i,j},v_{2,i,j},\cdots,v_{N_{\mathrm{RX}}^{\mathrm{RF}},i,j}]^\mathsf{T}$ can be regarded as the 2D DFT result of LS estimate $\hat{\mathbf{h}}_{\mathrm{rad},n,m}$ at occupied resource elements of an SSB. To enhance the output SNR, the energy across different receive RF chains can be accumulated. The delay-Doppler map after coherent combination is given by $\mathrm{DD}_{i,j} = |\sum_{p=1}^{N_{\mathrm{RX}}^{\mathrm{RF}}}v_{p,i,j}|^2.$
%\begin{equation}
%    \mathrm{DD}_{i,j} = |\sum_{p=1}^{N_{\mathrm{RX}}^{\mathrm{RF}}}v_{p,i,j}|^2.
%\end{equation}
The peaks of the delay-Doppler map yield the estimated delay $\hat{\tau}_l$, and Doppler shift $\hat{f}_{\mathrm{d},l}$, from where the range and radial velocity of the corresponding target can be calculated as $\hat{r}_l =  \frac{c\hat{\tau}_l}{2}$ and $\hat{v}_l = \frac{c \hat{f}_{\mathrm{d},l}}{2f_{\rm{c}}}$, respectively.
Since each SSB only contains 4 OFDM symbols, the velocity resolution is very low, i.e., two targets in nearby DD bins can not be resolved on the DD profile. To resolve this problem, we further explore the large antenna array to distinguish them in the angle domain. 

\subsection{Angular Estimation}
For each detected DD peak $(\hat{\tau}_l,\hat{f}_{\mathrm{d},l})$ with corresponding DFT bin index $(i_l,j_l)$, snapshots for spatial processing are gathered locally around the estimated range bin, i.e., $\{i_l-w_{\mathrm{r}},\cdots, i_l+w_{\mathrm{r}}\}$. $w_\mathrm{r}$ determines the range of the window. The snapshots for the $m$-th OFDM symbol are given by 
\begin{equation}
    \mathbf{Y}_l^{(m)} = [\mathbf{d}_{i_l-w_{\mathrm{r}},m},\cdots,\mathbf{d}_{i_l+w_{\mathrm{r}},m}].
\end{equation}
Stacking the snapshots for all $M$ OFDM symbols gives
\begin{equation}
    \mathbf{Y}_l = [\mathbf{Y}_l^{(1)},\cdots,\mathbf{Y}_l^{(M)}]\in \mathbb{C}^{N_{\mathrm{RX}}^{\mathrm{RF}} \times N_{\mathrm{ss}}},
\end{equation}
where $N_{\mathrm{ss}} = (2w_{\mathrm{r}}+1)M$ is the number of snapshots used for MUSIC algorithm.

The elevation angle and azimuth angle can be estimated through 2D augmented beamspace MUSIC. The beamspace MUSIC can effectively resolve the angular ambiguity caused by the hybrid beamforming structure. 
The spatial covariance for target $l$ is given by 
\begin{equation}
    \mathbf{R}_l = \frac{1}{N_{\mathrm{ss}}}\mathbf{Y}_l \mathbf{Y}_l ^{\mathsf{H}} \in \mathbb{C}^{N_{\mathrm{RX}}^{\mathrm{RF}}\times N_{\mathrm{RX}}^{\mathrm{RF}}}.
\end{equation}
The 2D argumented beamspace MUSIC spectrum can be calculated as
\begin{equation}
    P_l(\phi,\theta) = \frac{ \|\tilde{\mathbf{a}}(\phi,\theta)\|^2}{\tilde{\mathbf{a}}^\mathsf{H}(\phi,\theta) \Gamma_l \Gamma_l^{\mathsf{H}}\tilde{\mathbf{a}}(\phi,\theta)},
    \label{eq:music}
\end{equation}
where $\tilde{\mathbf{a}}(\phi,\theta) = (\mathbf{W}_{\mathrm{RX,rad}}^{\mathrm{RF}})^{\mathsf{H}} \mathbf{a}(\phi,\theta) \in \mathbb{C}^{N_{\mathrm{RX}}^{\mathrm{RF}} \times 1}$ is the beamspace steering vector, and $\Gamma_l$ is the matrix with eigenvectors of the noise subspace of $\mathbf{R}_l$. 
Unlike conventional MUSIC with the numerator being 1, the numerator in Eq. (\ref{eq:music}) is designed to be the normalization term to compensate for the non-uniform gain introduced by $\mathbf{W}_{\mathrm{RX,rad}}^{\mathrm{RF}}$.

The design of this sensing receive analog precoder $\mathbf{W}_{\mathrm{RX,rad}}^{\mathrm{RF}}$ affects both the detection performance in 2D DFT and the MUSIC performance. A DFT codebook-based beamforming can strictly enforce orthogonality between RF chains, i.e., $(\mathbf{W}_{\mathrm{RX,rad}}^{\mathrm{RF}})^\mathsf{H} \mathbf{W}_{\mathrm{RX,rad}}^{\mathrm{RF}} = \mathbf{I}$, which is desired in beamspace MUSIC algorithms. This analog precoder is thus designed with its columns selected from the DFT matrix. To further maximize the processing gain in 2D DFT peak detection, a locally-focused selection mechanism for the receive analog beamformer is designed. Specifically, the columns are selected to span the spatial subspace most correlated with the SSB sector. If we denote the DFT matrices as $\mathbf{D}_\mathrm{y} \in \mathbb{C}^{N_{\mathrm{RX,y}} \times N_{\mathrm{RX,y}}}$ and $\mathbf{D}_\mathrm{z} \in \mathbb{C}^{N_{\mathrm{RX,z}} \times N_{\mathrm{RX,z}}}$, the beamformer is designed as
\begin{equation}
\begin{split}
    \mathbf{W}_{\mathrm{RX,rad}}^{\mathrm{RF}} = \frac{1}{\sqrt{N_{\mathrm{RX}}}} \Big [ 
    &\mathbf{a}_{\mathrm{z}}^{(1)}\otimes \mathbf{a}_{\mathrm{y}}^{(1)}, \mathbf{a}_{\mathrm{z}}^{(1)}\otimes \mathbf{a}_{\mathrm{y}}^{(2)}, \cdots, \\
    &\mathbf{a}_{\mathrm{z}}^{(N_{\mathrm{RX,z}}^\mathrm{RF})}\otimes \mathbf{a}_{\mathrm{y}}^{(N_{\mathrm{RX,y}}^\mathrm{RF})} \Big],
\end{split}
\end{equation}
where ${a}_{\mathrm{z}}^{(p)}$ and ${a}_{\mathrm{y}}^{(q)}$ are columns from $\mathbf{D}_\mathrm{z}$ and $\mathbf{D}_\mathrm{y}$, so that $|(\mathbf{a}_{\mathrm{z}}^{(p)})^\mathsf{H}\mathbf{a}_\mathrm{z}(\phi_\mathrm{SSB})|$ gives the $p$-th largest result among all columns and $|(\mathbf{a}_{\mathrm{y}}^{(q)})^\mathsf{H}\mathbf{a}_\mathrm{y}(\theta_\mathrm{SSB})|$ gives the $q$-th largest result among all columns. $\phi_{\mathrm{SSB}}$ and $\theta_{\mathrm{SSB}}$ are the center direction of the current wide-beam SSB. This beamforming method not only enhances processing gain but also preserves the spatial diversity for the MUSIC algorithm.

The peak of the MUSIC spectrum provides the estimates of the elevation angle and azimuth angle $(\phi_l,\theta_l)$. 
Since each angle estimation is conditioned on a unique DD peak, the association between range-velocity and angles is inherently guaranteed. 
While target candidates are detected for every SSB, the sensing results are aggregated and associated by the BS to resolve the identities of the $K$ UAV users. This consolidated environmental awareness allows the BS to perform UAV tracking and user-specific robust beamforming in the subsequent communication phase.

\subsection{Computational Complexity}
Let \(B_{\mathrm{SSB}}\) denote the number of SSB beams processed in one SS burst, \(N_{\mathrm{IDFT}}\) and \(M_{\mathrm{DFT}}\) denote the delay- and Doppler-domain transform sizes, respectively, and \(N_{\mathrm{RX}}^{\mathrm{RF}}\) denote the number of receive RF chains. Furthermore, let \(N_{\mathrm{ss}}=(2w_r+1)M\) denote the number of local MUSIC snapshots and \(G_{\phi}\times G_{\theta}\) denote the two-dimensional angular-search grid.
The 2D-DFT processing and RF-chain energy combination over all SSB beams require
$\mathcal{O}\!\left(B_{\mathrm{SSB}}N_{\mathrm{RX}}^{\mathrm{RF}}
N_{\mathrm{IDFT}}M_{\mathrm{DFT}}
\log(N_{\mathrm{IDFT}}M_{\mathrm{DFT}})\right)$ operations.
For the $K$ detected UAVs, covariance construction, eigendecomposition,
and the MUSIC grid search require
$\mathcal{O}\!\left(K\left[
N_{\mathrm{ss}}(N_{\mathrm{RX}}^{\mathrm{RF}})^2+
(N_{\mathrm{RX}}^{\mathrm{RF}})^3+
G_{\phi}G_{\theta}(N_{\mathrm{RX}}^{\mathrm{RF}})^2
\right]\right)$ operations. Therefore, the overall computational complexity is
$
\mathcal{C}_{\mathrm{sense}}
=\mathcal{O}\Big(B_{\mathrm{SSB}}N_{\mathrm{RX}}^{\mathrm{RF}}
N_{\mathrm{IDFT}}M_{\mathrm{DFT}}
\log(N_{\mathrm{IDFT}}M_{\mathrm{DFT}})
+K\big[
N_{\mathrm{ss}}(N_{\mathrm{RX}}^{\mathrm{RF}})^2
+(N_{\mathrm{RX}}^{\mathrm{RF}})^3
+G_{\phi}G_{\theta}(N_{\mathrm{RX}}^{\mathrm{RF}})^2
\big]\Big).
$
These sensing operations are non-iterative and are performed only once per SSB burst. Moreover, the covariance construction and eigen-decomposition involve only small-dimensional matrices, while the MUSIC spectrum is evaluated only for the \(K\) detected range–velocity peaks. Therefore, the complexity of the sensing processing is relatively low compared with that of the subsequent robust beamforming optimization.

\section{EKF Tracking Procedure} \label{sec:tracking}
%The estimated UAV parameters are fed into EKFs for dynamic tracking between SSB bursts. Quantitative uncertainty measures can be obtained during EKF propagation, enabling robust characterization of channel parameters for beamforming optimization.

\subsection{Target Tracking and Prediction}
Since SSB-based sensing suffers from long periodicity and the UAVs may move rapidly, additional tracking mechanisms are required to maintain continuous acquisition of the channel parameters. To address the inherent nonlinearities in the motion dynamics and measurement models, an EKF is adopted.  

%By considering all targets estimated in different SSBs, 
In the designed EKF, the measurement vector for the $k$-th UAV at the $i$-th EKF iteration is given by $\mathbf{m}_k^{(i)} = [r_k^{(i)}, v_k^{(i)}, \phi_k^{(i)}, \theta_k^{(i)}]^\mathsf{T}$.
%\begin{equation}
%    \mathbf{m}_k^{(i)} = [r_k^{(i)}, v_k^{(i)}, \phi_k^{(i)}, \theta_k^{(i)}]^\mathsf{T}.
%\end{equation}
The state vector is designed as $\mathbf{s}_k^{(i)} = [x_k^{(i)},y_k^{(i)},z_k^{(i)}, v_{\mathrm{x},k}^{(i)},v_{\mathrm{y},k}^{(i)},v_{\mathrm{z},k}^{(i)}]^\mathsf{T},$
%\begin{equation}
%    \mathbf{s}_k^{(i)} = [x_k^{(i)},y_k^{(i)},z_k^{(i)}, v_{\mathrm{x},k}^{(i)},v_{\mathrm{y},k}^{(i)},v_{\mathrm{z},k}^{(i)}]^\mathsf{T},
%\end{equation}
where ($x_k^{(i)},y_k^{(i)},z_k^{(i)}$) denotes the position of the UAV in 3D coordinates, and ($v_{\mathrm{x},k}^{(i)},v_{\mathrm{y},k}^{(i)},v_{\mathrm{z},k}^{(i)}$) denotes the velocity component along the three axes. Assuming constant velocity over a short time period for UAV movement, we adopt a nearly constant velocity (NCV) model. The NCV model is not intended to impose a strict constant-velocity constraint; instead, deviations caused by acceleration and maneuvering are captured through the process noise and reflected in the prediction covariance. The state transition at the $i$-th iteration for the $k$-th UAV can be modeled as 
\begin{equation}
    \mathbf{s}_k^{(i+1)} = \mathbf{F} \mathbf{s}_k^{(i)} + \mathbf{u}_k^{(i)},
\end{equation}
where 
\begin{equation}
    \mathbf{F} = 
    \begin{bmatrix}
    \mathbf{I}_3&T_\mathrm{KF}\cdot\mathbf{I}_3\\
    \bm{0}_3&\mathbf{I}_3
    \end{bmatrix}
\end{equation}
is the transition matrix and $\mathbf{u}_k^{(i)}$ is the process noise vector, and $T_\mathrm{KF}$ is the duration time of each iteration. It is first assumed that $\mathbf{u}_k^{(i)}$ has zero mean and covariance matrix $\mathbf{Q}$. 
The potential for maneuvers in real-world UAV trajectories necessitates the inclusion of acceleration-induced uncertainties. To account for these deviations, a discrete white noise acceleration model is integrated for $\mathbf{Q}$ \cite{bar2001estimation}. Consequently, the process noise covariance matrix is formulated as
\begin{equation}
    \mathbf{Q} = \sigma_{\mathrm{a}}^2    \begin{bmatrix}
    \frac{1}{3}T_\mathrm{KF}^3\cdot\mathbf{I}_3&\frac{1}{2}T_\mathrm{KF}^2\cdot\mathbf{I}_3\\
    \frac{1}{2}T_\mathrm{KF}^2 \cdot \mathbf{I}_3&T_\mathrm{KF}\cdot\mathbf{I}_3
    \end{bmatrix},
    \label{eq:qbaseline}
\end{equation}
where $\sigma_{\mathrm{a}}^2$ is the acceleration variance with a suggested practical range of $0.5 \,a_\mathrm{M} \leq \sigma_\mathrm{a} \leq a_\mathrm{M}$, and $a_\mathrm{M}$ is the maximum acceleration magnitude \cite{bar2001estimation}. 

The state vector and the measurement vector can be associated with
\begin{equation}
    \mathbf{m}_k^{(i)} = \mathbf{g}(\mathbf{s}_k^{(i)}) + \boldsymbol{\mu}_k^{(i)},
\end{equation}
where the transformation function is given by
\begin{equation}
    \mathbf{g}(\mathbf{s}_k^{(i)})=
    \begin{pmatrix}
        \sqrt{\left(x_k^{(i)}\right)^2 + \left(y_k^{(i)}\right)^2 + \left(z_k^{(i)}\right)^2}\\
        \frac{x_k^{(i)} v_{\mathrm{x},k}^{(i)} + y_k^{(i)} v_{\mathrm{y},k}^{(i)} + z_k^{(i)} v_{\mathrm{z},k}^{(i)}}{\sqrt{\left(x_k^{(i)}\right)^2 + \left(y_k^{(i)}\right)^2 + \left(z_k^{(i)}\right)^2}} \\
         \mathrm{arcsin}\left(\frac{z_k^{(i)}}{\sqrt{\left(x_k^{(i)}\right)^2 + \left(y_k^{(i)}\right)^2 + \left(z_k^{(i)}\right)^2}}\right)\\
        \mathrm{arctan2}(y_k^{(i)},x_k^{(i)})
    \end{pmatrix},
\end{equation}
and $\boldsymbol{\mu}_k^{(i)}$ is the measurement noise with zero mean and covariance matrix $\mathbf{Q}' = \mathrm{diag}(\sigma_{\rm{r}}^2,\sigma_{\rm{v}}^2,\sigma_{\rm{\phi}}^2,\sigma_{\rm{\theta}}^2)$. The entries of the covariance matrix can be inferred from the sensing estimation errors, which are closely related to the adopted sensing signal processing skills. %, along with the sensing SNR.

To linearize the transformation between the state vector and the measurement vector, the Jacobian matrix for $\mathbf{g}$ is given by
\begin{equation}
    \begin{aligned}
        \mathbf{G} &= \frac{\partial\mathbf{g}(\mathbf{s})}{\partial\mathbf{s}} \\
        & = 
        \begin{pmatrix}
            \frac{x}{r} & \frac{y}{r} & \frac{z}{r} & 0 & 0 & 0 \\
            \frac{v_{\rm{x}}}{r} - \frac{tx}{r^3} & \frac{v_{\rm{y}}}{r} - \frac{ty}{r^3} & \frac{v_{\rm{z}}}{r} - \frac{tz}{r^3} & 
            \frac{x}{r} & \frac{y}{r} & \frac{z}{r} \\
            -\frac{zx}{r^2\sqrt{r^2-z^2}} & -\frac{zy}{r^2\sqrt{r^2-z^2}} & \frac{\sqrt{r^2-z^2}}{r^2} &
            0 & 0 & 0 \\
            -\frac{y}{x^2+y^2} & \frac{x}{x^2+y^2} & 0& 0 & 0 & 0 
        \end{pmatrix},
    \end{aligned}
\end{equation}
where $t = xv_{\rm{x}} + yv_{\rm{y}} + zv_{\rm{z}}$ and $r = \sqrt{x^2 + y^2 + z^2}$.

The procedure of EKF for the $k$-th UAV is summarized as follows:
\begin{enumerate}
    \item State prediction: 
    \begin{equation} \label{eq:prediction}
        \hat{\mathbf{s}}_k^{(i|i-1)} = \mathbf{F} \hat{\mathbf{s}}_k^{(i-1|i-1)}.
    \end{equation}
    \item Covariance matrix prediction:
    \begin{equation}
        \mathbf{P}_k^{(i|i-1)} = \mathbf{F} \mathbf{P}_k^{(i-1|i-1)} \mathbf{F}^\mathsf{T} + \mathbf{Q},
    \end{equation}
    where 
    %$\mathbf{Q}_{k-1}$ is the process noise covariance, which is adaptively scaled based on the innovation sequence. 
    $\mathbf{P}_k^{(i|i-1)}$ and $\mathbf{P}_k^{(i-1|i-1)}$ denote the predicted and updated state covariance matrix.
    \item Kalman gain calculation:
    \begin{equation}
        \mathbf{K}_k^{(i)} = \mathbf{P}_k^{(i|i-1)}  (\mathbf{G}_k^{(i)})^\mathsf{T}[\mathbf{G}_k^{(i)} \mathbf{P}_k^{(i|i-1)} (\mathbf{G}_k^{(i)})^\mathsf{T} + \mathbf{Q'}]^{-1},
    \end{equation}
    where $\mathbf{K}_k^{(i)}$ is the Kalman gain, and $\mathbf{G}_k^{(i)}$ can be obtained from $\mathbf{G}_k^{(i)} = \frac{\partial \mathbf{g}(\mathbf{s})}{\partial \mathbf{s}}|_{\mathbf{s} = \hat{\mathbf{s}}_k^{(i|i-1)}}$.
    \item State estimation update:
    \begin{equation}\label{eq:update}
        \hat{\mathbf{s}}_k^{(i|i)} = \hat{\mathbf{s}}_k^{(i|i-1)} + \mathbf{K}_k^{(i)}[\mathbf{m}_k^{(i)} - \mathbf{g}(\hat{\mathbf{s}}_k^{(i|i-1)})].
    \end{equation}
    \item Covariance update:
    \begin{equation}
        \mathbf{P}_k^{(i|i)} = (\mathbf{I} - \mathbf{K}_k^{(i)} \mathbf{G}_k^{(i)})\mathbf{P}_k^{(i|i-1)}.
    \end{equation}
\end{enumerate}

It should be noted that the state estimation update in Eq. (\ref{eq:update}) only executes at the time of the SS burst when there exists a measurement result. 
%In other words, the iteration time for the EKF is determined by the SS burst interval $T$. 
Given the long SS burst interval $T$ and the need for frequent beamformer updates for fast-moving UAVs, predicted states are required in the interim to support continuous beamforming design. Therefore, between two SS bursts, we can only get a predicted state according to Eq. (\ref{eq:prediction}), where $T_\mathrm{KF}$ is determined by the predicted time step. To support robust beamforming design, we need to further analyze the uncertainty of the prediction, especially for angle and range parameters, which are employed in the following robust beamforming design.

\subsection{Uncertainties for Predicted Parameters}
During UAV maneuvers, the NCV model may fail to capture the structural deviation, resulting in a model mismatch that leads to an underestimation of the true tracking error \cite{KFmse}. Therefore, the process noise term $\mathbf{u}_k^{(i)}$ should be treated as a disturbance term.
To address this, a posterior uncertainty correction term is introduced to characterize the uncertainty caused by model mismatch, which is quantified by 
\begin{equation}
    \Delta \mathbf{S}_k^{(i)} = \mathbf{S}_{k,\mathrm{emp}}^{(i)} - \mathbf{S}_{k,\mathrm{th}}^{(i)},
\end{equation}
where $\mathbf{S}_{k,\mathrm{emp}}^{(i)}$ and $\mathbf{S}_{k,\mathrm{th}}^{(i)}$ stand for the empirical and theoretical innovation covariance. The former can be calculated from the innovation at the $i$-th iteration, which is given by
\begin{equation}
    \boldsymbol{\nu}_k^{(i)} = \mathbf{m}_k^{(i)} - \mathbf{g}(\hat{\mathbf{s}}_k^{(i|i-1)}).
\end{equation}
The empirical innovation covariance is given by 
\begin{equation}
    \mathbf{S}_{k,\mathrm{emp}}^{(i)} = \frac{1}{J}\sum_{j=1}^J\boldsymbol{\nu}_k^{(i-j)} (\boldsymbol{\nu}_k^{(i-j)})^\mathsf{T},
\end{equation}
where $J$ is the size of the moving window of the innovation covariance.
Meanwhile, the theoretical covariance of $\boldsymbol{\nu}_k^{(i)}$ assumed by the EKF is given by
\begin{equation}\label{eq:S-theo}
    \mathbf{S}_{k,\mathrm{th}}^{(i)} 
    = \mathbf{G}_k^{(i)} \mathbf{P}_k^{(i|i-1)} (\mathbf{G}_k^{(i)})^\mathsf{T} 
    + \mathbf{Q}'.
\end{equation}
To ensure that $\Delta \mathbf{S}_k^{(i)}$ is positive semidefinite (PSD), it can be projected onto the PSD cone by eigenvalue truncation as
\begin{equation}
    \Delta \tilde{\mathbf{S}}_k^{(i)} = \mathbf{V} \,\mathrm{diag}(\max\{\lambda_j,0\}) \, \mathbf{V}^\mathsf{T},
    \label{eq:psd_projection}
\end{equation}
where $\mathbf{V}$ and $\lambda_j$ are the eigen vector and eigenvalues of $\Delta \mathbf{S}_k^{(i)}$. The PSD projection ensures that the corrected covariance mismatch matrix $\Delta \mathbf{S}_k^{(i)}$ remains physically valid as a covariance representation. Since subtracting two covariance matrices may yield indefinite results, eigenvalue truncation removes non-physical negative components while preserving the dominant uncertainty directions. %This operation has negligible impact on beamforming performance and serves primarily as a numerical safeguard for algorithmic stability.

To relate this innovation-domain mismatch to state-domain uncertainty, the pseudoinverse of the measurement Jacobian is used, which is given by
\begin{equation}
    (\mathbf{G}_k^{(i)})^\dagger =
    (\mathbf{G}_k^{(i)})^\mathsf{T} \left[ \mathbf{G}_k^{(i)} (\mathbf{G}_k^{(i)})^\mathsf{T} \right]^{-1}.
\end{equation}
This yields the bias covariance term
\begin{equation}
    \mathbf{P}_{k,\mathrm{bias}}^{(i)}
    = (\mathbf{G}_k^{(i)})^\dagger  
      \Delta \tilde{\mathbf{S}}_k^{(i)}  ((\mathbf{G}_k^{(i)})^{\dagger})^{\mathsf{T}}.
\end{equation}

Finally, the corrected state estimation covariance is defined as
\begin{equation}
    \tilde{\mathbf{P}}_k^{(i)} 
    = \mathbf{P}_k^{(i|i)} + \mathbf{P}_{k,\mathrm{bias}}^{(i)}.
\end{equation}
This matrix incorporates not only the EKF's internal uncertainty, but also the uncertainty arising from model mismatch and unmodeled target maneuvers. 

Based on the corrected state estimation covariance, the variance of the range, azimuth angle, and elevation angle for the $k$-th target at the $i$-th iteration can be approximated through a first-order Taylor expansion as
\begin{align}\label{eq:appVar}
        & (\sigma_{\mathrm{r},k}^{(i)})^2 \approx \nabla g_1^\mathsf{T} \tilde{\mathbf{P}}_k^{(i)} \nabla g_1, \\ \label{eq:appVar2}
        & (\sigma_{\mathrm{\phi},k}^{(i)})^2 \approx \nabla g_3^\mathsf{T} \tilde{\mathbf{P}}_k^{(i)} \nabla g_3, \\
        \label{eq:appVar3}
        & (\sigma_{\mathrm{\theta},k}^{(i)})^2 \approx \nabla g_4^\mathsf{T} \tilde{\mathbf{P}}_k^{(i)} \nabla g_4,
\end{align}
where $\nabla g_j = \frac{\partial g_j(\mathbf{s})}{\partial \mathbf{s}}|_{\mathbf{s} = \hat{\mathbf{s}}_k^{(i|i)}}$ is the $j$-th row of matrix $\mathbf{G}$.

When applying the above uncertainty analysis to the prediction step, with an iteration time interval of $T_\mathrm{KF}$, the calculation in Eq. (\ref{eq:S-theo}) sets the value of $\mathbf{Q}'$ to $\mathbf{0}$. Thus, we can get the uncertainty of the predicted results.

\subsection{Computational Complexity}
Let $d_\text{s}$ and $d_\text{m}$ denote the dimensions of the state and
measurement vectors, respectively. For all $K$ UAVs, the state
prediction and covariance propagation have a computational complexity
of
$\mathcal{C}_{\mathrm{EKF,pred}}
=\mathcal{O}(K d_\text{s}^3)$.
At an SSB measurement period, the measurement update, innovation-covariance calculation, PSD projection, and covariance correction require
$\mathcal{C}_{\mathrm{EKF,meas}}
=\mathcal{O}\!\left(
K\left[
d_\text{s}^3+d_\text{m}^3+d_\text{s}^2d_\text{m}+d_\text{s}d_\text{m}^2+Jd_\text{m}^2
\right]\right)$.
Overall, the EKF prediction, measurement update, and covariance correction have relatively low computational complexity. Since the state and measurement dimensions are fixed and small, the EKF-related computations are lightweight and can be efficiently executed at the short prediction interval.
\section{Multi-User Robust Beamforming} \label{sec:beamforming}
Leveraging prior information from SSB-based sensing can guide communication beamforming by approximating the channel correlation matrix without CSI feedback. To be specific, by fusing SSB-derived channel parameters of the $K$ users with statistical channel models, the beamforming weights $\mathbf{W}_{\mathrm{TX,com}} = \mathbf{W}_{\mathrm{TX,com}}^{\mathrm{RF}} \mathbf{W}_{\mathrm{TX,com}}^{\mathrm{BB}}$ can be designed proactively, reducing overhead and accelerating beam alignment, especially in high-mobility scenarios. In the following discussion, the subscript ``TX,com'' will be omitted for simplicity. Since the beamforming is available for every EKF iteration, the superscript ``(i)'' will also be omitted.

\subsection{Robust Model for Communications}
SINR quantifies the signal quality of a user in the presence of MUI and noise. The SINR at the $k$-th user is expressed as 
\begin{equation}
        \text{SINR}_k = \frac{|\mathbf{h}_k^{\mathsf{H}}\mathbf{w}_k|^2}{|\mathbf{h}_k^{\mathsf{H}}\sum_{i=1,i\neq k}^K \mathbf{w}_i|^2 + \sigma_{\mathrm{z},k}^2}.
\end{equation}

As the exact channel frequency response is intractable in predictive beamforming, to achieve robust communication, an average SINR dependent on the covariance matrix can be considered instead of the instantaneous SINR. The average SINR calculates the expectation within a short time interval, and is given by 
\begin{equation}
        \overline{\text{SINR}}_k = \frac{\mathbf{w}_k^{\mathsf{H}} \mathbb{E}[\mathbf{h}_k \mathbf{h}_k^{\mathsf{H}}] \mathbf{w}_k}{\sum_{i=1,i\neq k}^K \mathbf{w}_i^\mathsf{H} \mathbb{E}[\mathbf{h}_k \mathbf{h}_k^{\mathsf{H}}] \mathbf{w}_i + \sigma_{\mathrm{z},k}^2}.
\end{equation}
The channel correlation of the $k$-th user is given by $\mathbf{R}_k=\mathbb{E}[\mathbf{h}_k \mathbf{h}_k^{\mathsf{H}}]$. 
According to our channel model, which depends on the channel attenuation coefficient and steering vector, the channel correlation matrix can thus be given by %The construction of this correlation matrix can utilize the sensing information obtained from Phase 1. While the angle estimates can be used in steering vectors, given the complex path gain $\alpha_k = \sqrt{\beta_k}e^{j\varphi_k}$, where $\beta_k$ is the attenuation factor related to the relative distance between the BS and the UAV, and $\varphi_k$ denotes the phase shift caused by time and frequency selectivity, the range estimates can also be utilized. 
\begin{equation}
    \begin{aligned}
           \mathbf{R}_k&=\mathbb{E}[\mathbf{h}_k \mathbf{h}_k^{\mathsf{H}}]\\
           & = \mathbb{E}[\beta_k \mathbf{a}(\phi_k,\theta_k) \mathbf{a}^{\mathsf{H}}(\phi_k,\theta_k)] \\
           & = \iiint p_{\bm{r},\bm{\phi},\bm{\theta},k}(r,\theta,\phi) \beta(r) \mathbf{a}(\phi,\theta) \mathbf{a}^{\mathsf{H}}(\phi,\theta) \text{d}r \text{d}\phi \text{d}\theta, \\
    \end{aligned}
\end{equation}
where $\beta(r) = \frac{\lambda^2}{(4\pi r)^2}$ is the free-space path loss factor, and $p_{\bm{r},\bm{\phi},\bm{\theta},k}(r,\theta,\phi)$ is the joint probability distribution function of range, azimuth angle, and elevation angle. Due to the independence of channel parameters between attenuation and angle, the correlation matrix can be written as
\begin{equation}
    \mathbf{R}_k = C_{\mathrm{r},k} \cdot \mathbf{R}_{\mathrm{\phi},\mathrm{\theta},k},
\end{equation}
where 
\begin{equation}
    C_{\mathrm{r},k} = \int p_{\bm{r},k}(r)\beta_k(r) \text{d}r,
\end{equation}
\begin{equation}
    \mathbf{R}_{\mathrm{\phi},\mathrm{\theta},k} = \iint  p_{\bm{\phi},k}(\phi) p_{\bm{\theta},k}(\theta) \mathbf{a}(\phi,\theta) \mathbf{a}^{\mathsf{H}}(\phi,\theta) \text{d} \phi \text{d} \theta,
\end{equation}
are the correlation constant for attenuation, and the correlation matrix for angles, respectively. $p_{\bm{r},k}(r)$, $p_{\bm{\phi},k}(\phi)$, and $p_{\bm{\theta},k}(\theta)$ are the probability density functions (PDFs) of range, elevation angle, and azimuth angle, respectively. 

Given the approximated variances of channel parameters in Eqs. (\ref{eq:appVar}), (\ref{eq:appVar2}) and (\ref{eq:appVar3}), a robust assumption about the distribution of target parameters can be established by considering the uncertainty of the EKF. Specifically, assuming that the estimation errors are zero-mean and Gaussian-distributed, the PDF for each parameter can be derived as
\begin{align}
    & p_{\bm{r},k}(r) = \frac{1}{\sqrt{2 \pi (\sigma_{\mathrm{r},k})^2}} \mathrm{exp}\left[-\frac{(r-\hat{r}_k)^2}{2(\sigma_{\mathrm{r},k})^2} \right], \\
    & p_{\boldsymbol{\phi},k}(\phi) = \frac{1}{\sqrt{2 \pi (\sigma_{\mathrm{\phi},k})^2}} \mathrm{exp}\left[-\frac{(\phi-\hat{\phi}_k)^2}{2(\sigma_{\mathrm{\phi},k})^2} \right], \\
    & p_{\boldsymbol{\theta},k}(\theta) = \frac{1}{\sqrt{2 \pi (\sigma_{\mathrm{\theta},k})^2}} \mathrm{exp}\left[-\frac{(\theta-\hat{\theta}_k)^2}{2(\sigma_{\mathrm{\theta},k})^2} \right].
\end{align}
%where $\hat{r}_k^{(i)}$, $\hat{\phi}_k^{(i)}$, $\hat{\theta}_k^{(i)}$ are the prediction results for the $k$-th target at the $i$-th iteration.
These PDFs quantify the uncertainty in the estimation due to measurement, process noise, and the model mismatch, aiding robust communication beamforming. 

With the robust PDFs for range and angles, the channel correlation matrix can be calculated. To avoid extremely small values in subsequent optimization, an effective noise variance is defined for the $k$-th user as $\sigma_{\mathrm{eff},k}^2 = \frac{\sigma_{\mathrm{z},k}^2}{C_{\mathrm{r},k}}$, and the corresponding average SINR can be rewritten as 
\begin{equation}
    \overline{\text{SINR}}_k = \frac{\mathbf{w}_k^{\mathsf{H}} \mathbf{R}_{\mathrm{\phi},\mathrm{\theta},k} \mathbf{w}_k}{\sum_{i=1,i\neq k}^K \mathbf{w}_i^\mathsf{H} \mathbf{R}_{\mathrm{\phi},\mathrm{\theta},k} \mathbf{w}_i + \sigma_{\mathrm{eff},k}^2}.
\end{equation}

In multi-user scenarios, the sum-rate measures the system’s overall capacity. Maximizing the sum-rate through coordinated beamforming ensures efficient spectrum utilization. %Considering the insufficient knowledge of CSI, the average sum rate calculated based on the average SINR of different users can be regarded as the communication metric within a short time interval. 
The resulting average sum-rate for the entire network is given by
\begin{equation}
    \overline{\rm{SR}} = \sum_{k=1}^K \text{log}_2 (1+\overline{\text{SINR}}_k).
\end{equation}

\subsection{Problem Formulation}
To maximize the sum-rate of the entire network while guaranteeing stable connections between BS and UAVs, the optimization problem can be formulated as follows.
\begin{subequations}
\begin{align}
\underset{\mathbf{W}}{\max} \quad  & \overline{\rm{SR}} \label{eq:objective}\\
\text {s.t. } &  \overline{\text{SINR}}_k \geq \gamma_k,\quad \forall k\label{eq:SINRConstraint}\\
& \| \mathbf{W}\|_\mathbf{F}^2 \leq P_\text{max}\label{eq:powerConstraint} \\
&     |(\mathbf{W}_{\rm{TX}}^{\rm{RF}})_{i,j}| = \!\frac{1}{\sqrt{N_{\rm{TX}}}},\forall i \in [1,N_{\rm{TX}}], j\in [1,N_{\rm{TX}}^{\rm{RF}}]\label{eq:WRFConstraint},
\end{align}
\end{subequations}
where $\gamma_k$ is the SINR threshold for each user $k$. This formulation aims at maximizing the average sum-rate, with constraint (\ref{eq:SINRConstraint}) ensuring the link quality for individual users, constraint (\ref{eq:powerConstraint}) limiting the total transmit power, and constraint (\ref{eq:WRFConstraint}) complying with the hybrid beamforming structure.

\subsection{Solution Algorithm}
The non-convexity of the objective function and SINR constraints, as well as the constant envelope constraint due to the presence of the analog beamformer, make the proposed optimization problem intractable at first glance. To simplify the optimization, the problem can be first approached by treating the beamforming structure as fully digital, which removes the constraint (\ref{eq:WRFConstraint}).

To this end, denoting $\mathbf{W}_k = \mathbf{w}_k \mathbf{w}_k^{\mathsf{H}} $, with $\mathbf{W}_k \succeq 0$ and $\mathrm{rank}(\mathbf{W}_k) = 1$, the formulation can be expressed as
\begin{subequations}
\begin{align}
\underset{\{\mathbf{W}_k \succeq 0\}}{\max} \quad  & \sum_{k=1}^K \mathrm{log}_2 \!\left(\!1\!+\!\frac{\mathrm{tr}(\mathbf{R}_{\rm{\phi,\theta},k} \mathbf{W}_k)}{\sum_{i \neq k}\mathrm{tr}(\mathbf{R}_{\mathrm{\phi,\theta},k}\mathbf{W}_i) + \sigma_{\mathrm{eff},k}^2}\!\right) \label{eq:objective2}\\
\text {s.t. } &  \mathrm{tr}(\mathbf{R}_{\mathrm{\phi,\theta},k} \mathbf{W}_k)\! \geq \!\gamma_k \!\left(\! \sum_{i \neq k}\mathrm{tr}(\mathbf{R}_{\mathrm{\phi,\theta},k}\mathbf{W}_i) \!+\! \sigma_{\mathrm{eff},k}^2 \!\right), \forall k\label{eq:SINRConstraint2}\\
& \sum_{k=1}^K \mathrm{tr}(\mathbf{W}_k)  \leq P_\text{max} \label{eq:powerConstraint2} \\
& \text{rank}(\mathbf{W}_k) = 1, \quad \forall k \label{eq:rankConstraint2}.
\end{align}
\end{subequations}

The semidefinite relaxation (SDR) framework can be applied, dropping the rank constraint, but the relaxed problem is still non-convex due to the objective function. To address the non-convexity of the objective function, an SCA approach is employed. The SCA algorithm begins with a feasible initial solution, and the convex form obtained from the first-order Taylor expansion of the non-convex objective is used alternatively in each iteration. Specifically, let 
\begin{align}
    & a_k = \mathrm{tr}(\mathbf{R}_{\rm{\phi,\theta},k} \mathbf{W}_k), \label{eq:xk} \\
    & b_k = \sum_{i \neq k}\mathrm{tr}(\mathbf{R}_{\mathrm{\phi,\theta},k}\mathbf{W}_i) + \sigma_{\mathrm{eff},k}^2. \label{eq:yk}
\end{align}
With $a_k,b_k > 0$, the $k$-th term in the objective has
\begin{equation}
    \mathrm{log}_2\left(1+\frac{a_k}{b_k}\right) = \mathrm{log}_2\left(a_k+b_k\right)-\mathrm{log}_2\left(b_k\right).
\end{equation}
The convexity characteristic of $-\mathrm{log}_2\left(b_k\right)$ gives
\begin{equation}
    -\mathrm{log}_2\left(b_k\right) \geq -\mathrm{log}_2\left(b_k^{(n)}\right) - \frac{1}{b_k^{(n)}{\ln2}}(b_k-b_k^{(n)}).
\end{equation}
Omitting the constant terms, the objective function at the $n$-th optimization iteration can be written as 
\begin{equation}
\begin{aligned}
    \tilde{\mathrm{SR}}^{(n)} =&  \sum_{k=1}^K  \bigg[\mathrm{log}_2\left(\sum_{j }\mathrm{tr}(\mathbf{R}_{\mathrm{\phi,\theta},k}\mathbf{W}_j)  + \sigma_{\mathrm{eff},k}^2\right) \\
    & \quad\quad\quad-\frac{1}{b_k^{(n)}}\left(\sum_{j \neq k}\mathrm{tr}(\mathbf{R}_{\mathrm{\phi,\theta},k}\mathbf{W}_j) - \sigma_{\mathrm{eff},k}^2\right)\bigg].
\end{aligned}
\end{equation}
The convex problem at the $n$-th iteration of SCA is given by
\begin{equation}
\begin{aligned}
\underset{\{\mathbf{W}_k \succeq 0\}}{\max} \quad  &\tilde{\mathrm{SR}}^{(n)} \\
\text {s.t. } & (\ref{eq:SINRConstraint2})\\
& (\ref{eq:powerConstraint2}),
\end{aligned}
\label{eq:subProblem}
\end{equation}
where $b_k^{(n)}$ is the optimal solution of $b_k$ obtained in the $n$-th iteration. 
To obtain a feasible initial point, we solve the following minimum-power problem over the SDR-relaxed domain:
\begin{equation}
\begin{aligned}
P_{\mathrm{init}}
=\min_{\{\mathbf{W}_k\succeq\mathbf{0}\}_{k=1}^{K}}
&\quad \sum_{k=1}^{K}\operatorname{tr}(\mathbf{W}_k)\\
\mathrm{s.t.}\quad
&a_k\geq\gamma_k b_k,
\quad \forall k .
\end{aligned}
\label{eq:feasible_initialization}
\end{equation}
If  problem \eqref{eq:feasible_initialization} is infeasible or
$P_{\mathrm{init}}>P_{\max}$, the prescribed QoS requirements
are infeasible under the available power budget. Otherwise, its
optimizer satisfies \eqref{eq:SINRConstraint2}--\eqref{eq:powerConstraint2} and is
adopted as the initial point $\mathbf{W}_k^{(0)}$.
The SCA solution can thus be obtained following Algorithm \ref{alg:sca_solution}.
Since each SCA update maximizes a tight global lower bound over the same feasible set, the original objective value is monotonically nondecreasing. The feasible set is compact, and the objective is bounded; hence, the objective-value sequence converges. Under the standard SCA regularity conditions, every accumulation point satisfies the KKT conditions of the SDR-relaxed problem.

\begin{algorithm}[t]
\caption{SDR-SCA for Fully Digital Beamformer Optimization}
\label{alg:sca_solution}
\begin{algorithmic}[1]
\STATE \textbf{Input:} $\{\mathbf{R}_{\phi,\theta,k}\}$, $\{\sigma_{\mathrm{eff},k}^2\}$, $\{\gamma_k\}$, $P_{\max}$.
\STATE \textbf{Output:} Optimal fully digital covariance $\{\mathbf{W}_k^*\}$.
\STATE Set $i=0$
\STATE Set convergence threshold $\epsilon$. 
%\STATE Find a feasible starting solution $\{\mathbf{W}_k^{(0)}\}$.
\STATE Obtain $\{\mathbf{W}_k^{(0)}\}$ by solving
\eqref{eq:feasible_initialization}.
\STATE Calculate $\{b_k^{(0)}\}$ based on $\{\mathbf{W}_k^{(0)}\}$ for each $k$ according to (\ref{eq:yk}).
\REPEAT
    \STATE $i \leftarrow i+1$.
    \STATE Solve the convex SDP problem ($\ref{eq:subProblem}$) to obtain $\{\mathbf{W}_k^{(i)}\}$.
    \STATE Calculate and $\{b_k^{(i)}\}$ using $\{\mathbf{W}_k^{(i)}\}$ according to (\ref{eq:yk}).
\UNTIL {$|\tilde{\mathrm{SR}}^{(i)} - \tilde{\mathrm{SR}}^{(i-1)}| \le \epsilon$ .}
\end{algorithmic}
\end{algorithm}

After the SCA method, randomization is needed to approach the rank-1 constraints. Apply SVD on the SCA results $\{\mathbf{W}^*_k\}$ as $\mathbf{W}_k^* = \mathbf{U}_k\mathbf{\Sigma}_k\mathbf{U}_k^\mathsf{H}$ and denote $\mathbf{L}_k = \mathbf{U}_k\mathbf{\Sigma}_k^{1/2}$. The following steps are repeated to select an optimal solution: 1) For each $k$, let $\mathbf{z}_k \sim \mathcal{CN}(\bm{0},\mathbf{I})$. Let $\mathbf{w}_k' = \mathbf{L}_k\mathbf{z}_k$; 2) Check whether the obtained $\mathbf{w}_k'$ satisfy the given constraints. If yes, calculate the corresponding average sum-rate.
%\begin{enumerate}
%    \item For each $k$, let $\mathbf{z}_k \sim \mathcal{CN}(\bm{0},\mathbf{I})$. Let $\mathbf{w}_k' = \mathbf{L}_k\mathbf{z}_k$.
%    \item Check whether the obtained $\mathbf{w}_k'$ satisfy the given constraints. If yes, calculate the corresponding average sum-rate.
%\end{enumerate}
The maximum average sum-rate can be obtained with $\mathbf{W}^* = [\mathbf{w}_1^*,\mathbf{w}_2^*,\cdots,\mathbf{w}_K^*]$.

\begin{figure*}[t]
    \centering
    \subfigure[Range-Velocity map]
    {\includegraphics[width = 0.3\linewidth]{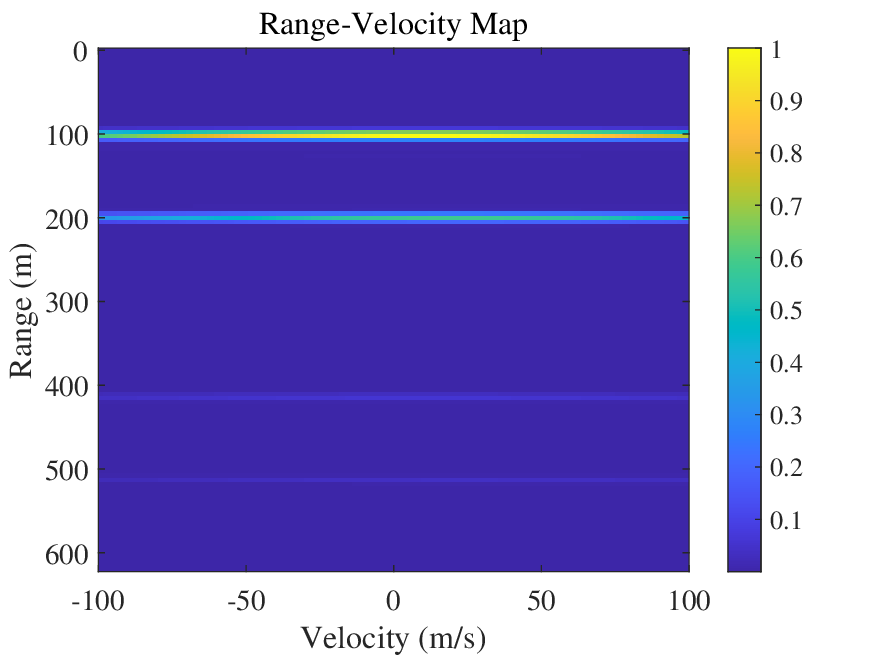}\label{fig:rvmap}}
    \subfigure[MUSIC spectrum for range-velocity peak 1]
    {\includegraphics[width=0.3\textwidth]{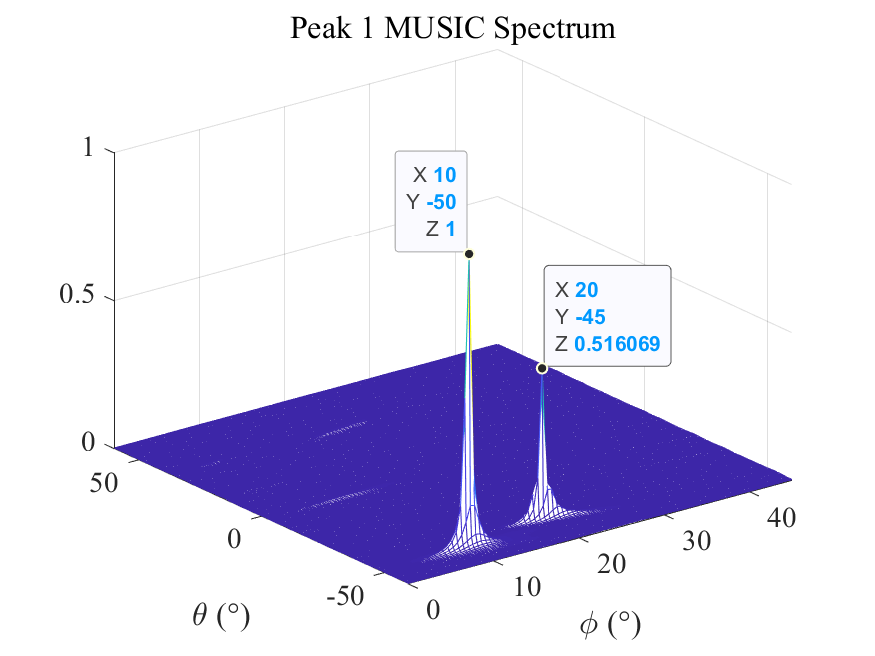}\label{fig:target1music}}
    \subfigure[MUSIC spectrum for range-velocity peak 2]
    {\includegraphics[width=0.3\linewidth]{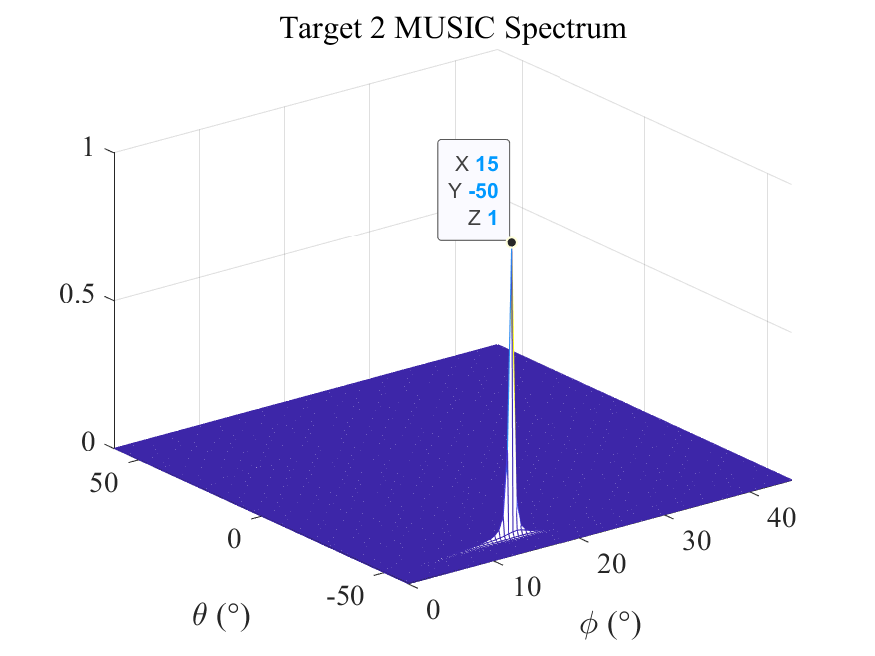}\label{fig:target2music}}
    \caption{Sensing results for the three targets within the coverage of one SSB beam.
}\label{fig:sensing}
    \vspace{-1em}
\end{figure*}

Given the optimal fully-digital transmit beamformer $\mathbf{W}^*$, the analog and digital beamformers for the fully connected hybrid beamforming structure can be approached by the following optimization problem.
\begin{equation}
    \begin{aligned}
        \underset{\mathbf{W}_{\rm{TX}}^{\rm{RF}},\mathbf{W}_{\rm{TX}}^{\rm{BB}}}{\min} \quad  & \|\mathbf{W}^*-\mathbf{W}_{\rm{TX}}^{\rm{RF}}\mathbf{W}_{\rm{TX}}^{\rm{BB}}\|_{\mathbf{F}}\\
\text {s.t. } &      |(\mathbf{W}_{\rm{TX}}^{\rm{RF}})_{i,j}|\! =\!\! \frac{1}{\sqrt{N_{\rm{TX}}}},\forall i \!\in \![1,N_{\rm{TX}}], j\!\in\! [1,N_{\rm{TX}}^{\rm{RF}}]\\
& \|\mathbf{W}_{\rm{TX}}^{\rm{RF}}\mathbf{W}_{\rm{TX}}^{\rm{BB}}\|_2^2  \leq P_\text{max}.
    \end{aligned}
    \label{eq:hb_problem}
\end{equation}
Alternating minimization methods can be adopted to approach the fully-digital beamformer with analog and digital beamformers, as given in \cite{altermini}. Specifically, the digital beamformer and analog beamformer can be alternately optimized to obtain the hybrid form. A solution is accepted only if the randomized fully digital and final hybrid beamformers satisfy their respective original constraints within the prescribed numerical tolerances; thus, the feasibility guarantee is conditional on these explicit checks.

\subsection{Computational Complexity}
If the angular integrals in the channel correlation matrices are evaluated using
$Q_{\phi}\times Q_{\theta}$ quadrature points, their construction requires $C_{\mathrm{cov}}=\mathcal{O}\left(KQ_{\phi}Q_{\theta}N_{\mathrm{TX}}^{2}\right)$.
Let $I_{\mathrm{SCA}}$, $I_{\mathrm{IP}}$, $N_{\mathrm{GR}}$, and
$I_{\mathrm{AM}}$ denote the numbers of SCA iterations, interior-point
iterations, Gaussian-randomization trials, and alternating-minimization
iterations, respectively. After SDR, the optimization contains $K$ Hermitian
PSD matrices of size $N_{\mathrm{TX}}\times N_{\mathrm{TX}}$. For a block-structured primal-dual solver, the computational complexity of the initialization and SDR-SCA stages can be conservatively characterized as
$
C_{\mathrm{sub}}
=
\mathcal{O}\left(
I_{\mathrm{IP}}
\left[
KN_{\mathrm{TX}}^{3}
+m_c^{2}KN_{\mathrm{TX}}^{2}
+m_c^{3}
\right]
\right),
$
where $m_c=\mathcal{O}(K)$ is the number of scalar coupling constraints.
Including the feasible-initialization problem, the total SDR-SCA complexity is
$C_{\mathrm{init+SCA}}
=
\mathcal{O}\left((1+I_{\mathrm{SCA}})C_{\mathrm{sub}}\right)$.
The Gaussian-randomization and hybrid-approximation stages respectively require
$
C_{\mathrm{GR}}
=
\mathcal{O}\left(
KN_{\mathrm{TX}}^{3}
+
N_{\mathrm{GR}}
\left[
KN_{\mathrm{TX}}^{2}
+K^{2}N_{\mathrm{TX}}^{2}
+C_{\mathrm{LP}}(K)
\right]
\right),
$
and
$
C_{\mathrm{hyb}}
=\mathcal{O}\!\left(
I_{\mathrm{AM}}N_{\mathrm{TX}}N_{\mathrm{TX}}^{\mathrm{RF}}
\left(N_{\mathrm{TX}}^{\mathrm{RF}}+K\right)
\right)
$.
Therefore, the overall robust-beamforming complexity is
$C_{\mathrm{BF}}
=
C_{\mathrm{cov}}
+C_{\mathrm{init+SCA}}
+C_{\mathrm{GR}}
+C_{\mathrm{hyb}}$.
The SDR-SCA stage is the dominant computational component due to large PSD matrices. Nevertheless, the Gaussian-randomization trials are independent and can be processed in parallel. Thus, the computational burden is manageable for a BS.

\section{Simulation Results}\label{sec:sim}
The simulation setup of the OFDM system follows 3GPP standards \cite{38211}. The carrier frequency is $f_\mathrm{c} = 28$ GHz, and the subcarrier spacing is $\Delta_\mathrm{f} = 120$ kHz. The bandwidth is set to $B = 100$ MHz, and the symbol duration is given by $T_\mathrm{o} = 8.9\text{ }\mu$s. The number of antennas at the BS is assumed to be $N_{\mathrm{TX}} = N_{\mathrm{RX}} = 256$, with $N_{\mathrm{TX,y}} = N_{\mathrm{RX,y}} = 32$ and $N_{\mathrm{TX,z}} = N_{\mathrm{RX,z}} = 8$. The number of RF chains is assumed to be $N_{\mathrm{TX}}^{\mathrm{RF}} = N_{\mathrm{RX}}^{\mathrm{RF}} = 16$, with $N_{\mathrm{TX,y}}^{\mathrm{RF}} = N_{\mathrm{RX,y}}^{\mathrm{RF}} =N_{\mathrm{TX,z}}^{\mathrm{RF}} = N_{\mathrm{RX,z}}^{\mathrm{RF}}= 4$. 

\subsection{Sensing Performance}
% First, the performance of the SSB-based sensing algorithm is evaluated, where the number of DFT points adopted is $N_{\mathrm{IDFT}} = M_{\mathrm{DFT}} = 256$, and the MUSIC spectrum search is realized by a step size of $1^\circ$ and $w_\mathrm{r} = 2$. 
% In accordance with the 3GPP NR standard for FR2 (Frequency Range 2) \cite{38213}, a maximum of $L_\mathrm{max} = 64$ SSBs is configured. To provide seamless coverage over the region of interest, which spans $[-60^\circ, 60^\circ]$ in azimuth and $[0^\circ, 45^\circ]$ in elevation, the 64 available SSB beams are organized into a $16 \times 4$ grid. Consider a scenario involving three UAVs, with ranges of $[100,200,120]$ m, radial velocities of $[10,50,-15]$ m/s, azimuth angles of $[-50^\circ, -45^\circ, 40^\circ]$, and elevation angles of $[10^\circ,20^\circ,37.5^\circ]$. 
% To evaluate the ability of the sensing algorithm to distinguish and track targets within one SSB beam, we specifically focus on the first two UAVs. In this setup, these two UAVs are closely spaced within the spatial coverage of the same SSB beam. 
First, the effectiveness of the SSB-based sensing algorithm is evaluated, where the number of DFT points adopted is $N_{\mathrm{IDFT}} = M_{\mathrm{DFT}} = 256$, and the MUSIC spectrum search is realized by a step size of $1^\circ$ and $w_\mathrm{r} = 2$. 
In accordance with the 3GPP NR standard for FR2 (Frequency Range 2) \cite{38213}, a maximum of $64$ SSBs is configured for each SS burst. To provide seamless coverage over the region of interest, which spans $[-60^\circ, 60^\circ]$ in azimuth and $[0^\circ, 45^\circ]$ in elevation, the 64 available SSB beams are organized into a $16 \times 4$ grid. 
Consider a scenario involving three UAVs, with ranges of $[100,103,200]$ m, radial velocities of $[10,12,20]$ m/s, azimuth angles of $[-50^\circ, -45^\circ, -50^\circ]$, and elevation angles of $[10^\circ,20^\circ,15^\circ]$.
To evaluate the ability of the sensing algorithm to distinguish and track targets within one SSB beam, the three UAVs are set closely spaced within the spatial coverage of the same SSB beam, while the first two UAVs are indistinguishable from the range-velocity map.

% As shown in Fig. \ref{fig:sensing}, the UAVs can be successfully detected. The range-velocity map obtained from 2D DFT is given by Fig. \ref{fig:rvmap}. Due to the frame structure of SSB, the small number of OFDM symbols leads to insufficient velocity resolution. However, as long as the targets can be distinguished by range or angles, the targets can be resolved. The 2D MUSIC spectra for individual targets are provided in Fig. \ref{fig:target1music} and Fig. \ref{fig:target2music}, respectively. Beamspace MUSIC requires spatial orthogonality of the beamforming matrix to suppress interference and noise. Although the proposed receive beamformer does not meet perfect spatial orthogonality, the spectral leakage in the MUSIC spectra generally does not affect MUSIC detection. 
As shown in Fig. \ref{fig:sensing}, all UAVs can be successfully detected and estimated. The range-velocity map obtained from 2D DFT is given in Fig. \ref{fig:rvmap}. Due to the frame structure of SSB, the small number of OFDM symbols leads to insufficient velocity resolution. However, as long as the targets can be distinguished by range or angles, the targets can be resolved. The 2D beamspace MUSIC spectra for individual peaks are provided in Fig. \ref{fig:target1music} and Fig. \ref{fig:target2music}, respectively. As we can see, our proposed design yields sharp peaks in the MUSIC spectra. In Fig. \ref{fig:target1music}, the two targets that cannot be resolved in the range-velocity map are effectively resolved in the 2D angle domains. %Thus, all the targets can be effectively detected and estimated with our hierarchical signal processing design. 
%Beamspace MUSIC requires spatial orthogonality of the beamforming matrix to suppress interference and noise.

% \begin{figure}[t]
%     \centering
%     \includegraphics[width=0.4\textwidth]{Figures/CFAR_1231.eps}
%     \vspace{-1em}
%     \caption{2D FFT detection probability of two receive beamforming designs.}
%     \label{fig:cfar}
%     \vspace{-1em}
% \end{figure}

% The design of the sensing receive beamformer $\mathbf{W}_{\mathrm{RX,rad}}^{\mathrm{RF}}$ determines not only the pattern of the MUSIC spectra but also the detection performance in range-velocity profiling.
% The detection probability versus SNR for the proposed locally-focused method and the conventional DFT-based method is illustrated in Fig. \ref{fig:cfar}. For a fair comparison, the DFT-based receive beamformer weight is selected as the column from the DFT codebook that aligns most closely with the SSB transmit beam. The sensing detection is performed using a cell-average constant false alarm rate (CA-CFAR) detector with a target false alarm rate of $P_\mathrm{FA} = 10^{-5}$. As shown in Fig. \ref{fig:cfar}, the proposed method consistently outperforms the DFT-based approach, achieving an approximately 3 dB gain. 
% Since the proposed method accumulates energy within the SSB's region of interest, a decreased SNR threshold required for reliable detection can be achieved.

\subsection{Communication Performance}

% \begin{figure}[t]
%     \centering
%     \includegraphics[width=0.4\textwidth]{Figures/trajectory_0106.eps}
%     \caption{Trajectory and EKF results.}
%     \label{fig:trajectory}
% \end{figure}

%The communication rate performance of the proposed beamforming scheme is evaluated. 
The feedback-based beamforming method is adopted as the benchmark, which includes the SSB-based coarse beam selection and CSI-RS-based beam refinement/tracking with feedback. The CSI-RS-based beam refinement/tracking can improve the beam alignment accuracy, but leads to extra overhead for each user. In our evaluation, we consider two configurations for CSI-RS: 10 ms interval with 4 extra refinement beams/symbols and 5 ms interval with 64 refinement beams/symbols (the latter is marked with `denser' in the figures). 
%The feedback-based beamforming methods are adopted as benchmarks, where, besides the SSB beam selection procedure, uplink feedback for the SSB strength is required, and refinement is carried out for a fixed time interval, both of which can cause extra overhead. The denser feedback means that more angles are swept during beam refinement, and the refinement interval is shorter.
%Since existing predictive beamforming designs require prior knowledge of motion trajectories and do not consider the multi-user scenario, they are not applicable for comparison with our design.
Table~\ref{tab:overhead_ratio} compares the signaling overhead ratios of the three beamforming schemes under different SSB periods for $K=1$ and $K=3$. For $K=1$, as the SSB period increases from 10~ms to 160~ms, the overhead ratios of the proposed beamforming and feedback-based beamforming decrease from 50.00\% to 3.13\% and from 50.54\% to 3.58\%, respectively. However, the feedback-based beamforming (denser) incurs substantially higher overhead in the refinement process. More importantly, the signaling overhead of the proposed beamforming scheme is independent of $K$, since it requires neither user-specific CSI-RS beam refinement nor uplink feedback. In contrast, the overhead of both feedback-based schemes increases with $K$, particularly for the denser scheme, because the refinement resources scale with the number of UAVs.
\begin{table}[!t]
\centering
\caption{Overhead Ratios Under Different SSB Periods}
\label{tab:overhead_ratio}
\scriptsize
\renewcommand{\arraystretch}{0.92}
\setlength{\tabcolsep}{0.8pt}

\begin{tabularx}{\columnwidth}{
@{}
>{\centering\arraybackslash}p{0.35\columnwidth}
>{\centering\arraybackslash}p{0.045\columnwidth}
*{5}{>{\centering\arraybackslash}X}
@{}}
\toprule
\textbf{Scheme}
& \boldmath{$K$}
& \textbf{10 ms}
& \textbf{20 ms}
& \textbf{40 ms}
& \textbf{80 ms}
& \textbf{160 ms} \\
\midrule

\multirow{2}{*}{\shortstack[c]{Proposed beamforming}}
& 1 & 50.00\% & 25.00\% & 12.50\% & 6.25\% & 3.13\% \\
& 3 & 50.00\% & 25.00\% & 12.50\% & 6.25\% & 3.13\% \\
\addlinespace[1pt]

\multirow{2}{*}{\shortstack[c]{Feedback-based beamforming}}
& 1 & 50.54\% & 25.49\% & 12.97\% & 6.71\% & 3.58\% \\
& 3 & 51.61\% & 26.47\% & 13.91\% & 7.62\% & 4.48\% \\
\addlinespace[1pt]

\multirow{2}{*}{\shortstack[c]{Feedback-based\\beamforming (denser)}}
& 1 & 55.89\% & 33.75\% & 22.68\% & 17.14\% & 14.38\% \\
& 3 & 67.68\% & 51.25\% & 43.04\% & 38.93\% & 36.88\% \\
\bottomrule
\end{tabularx}
\vspace{-1em}
\end{table}

Then, we investigate a 3D tracking scenario involving a highly maneuvering UAV over a 4-second duration. The BS is set at the origin $(0,0,0)$, and the UPA is placed along the y-axis and z-axis. The UAV starts from $(50,30,8)$ m and follows a nonlinear trajectory through a sequence of predefined waypoints, eventually reaching $(42,25,10)$ m. The trajectory consists of an approximately linear initial segment, followed by a pronounced turning segment, and an approximately linear terminal segment. The total trajectory length is approximately $57.76$ m over a duration of $4$ s, corresponding to an average UAV speed of approximately $14.44$ m/s. This trajectory provides relatively smooth motion before and after the turning segment while introducing a substantial change in the UAV's moving direction during the maneuver, thereby providing a representative scenario for evaluating beam tracking under nonlinear UAV mobility. The EKF integrates  SSB sensing measurements ($T = 40$ ms) with high-frequency prediction 
%($T_\mathrm{KF} = 1$ ms)
($T_\mathrm{KF} = 5$ ms) to support communication beamforming.
%We investigate a 3D tracking scenario involving a highly maneuvering UAV over an 8-second duration. The BS is set at the origin (0,0,0), and the UPA is placed along the y-axis and z-axis. The UAV follows a nonlinear trajectory, including a constant initial velocity of $[15,10,8]$ m/s and an abrupt maneuvering interval with a 3D acceleration of $[5, -15, 5] \text{ m/s}^2$ during 3 s to 5 s. The EKF integrates periodic SSB sensing measurements ($T = 40$ ms) with high-frequency prediction ($T_\mathrm{KF} = 1$ ms) to support communication beamforming. In the trajectory visualization, the tracking uncertainty is quantified by $\sqrt{\text{tr}(\tilde{\mathbf{P}}_k)}$, depicted as red circles centered at the estimated positions. These circles represent the spatial confidence interval of the EKF, where the variation of the radius during the acceleration phase captures the possible prediction error caused by sudden maneuvers. 

% \begin{figure*}[t]
%     \centering
%     \subfigure[Beam pattern for user 1]
%     {\includegraphics[width = 0.32\linewidth]{Figures/user1beampattern_cut_1230.eps}\label{fig:u1bp}}
%     \subfigure[Beam pattern for user 2]
%     {\includegraphics[width=0.32\textwidth]{Figures/user2beampattern_cut_1230.eps}\label{fig:u2bp}}
%     \subfigure[Beam pattern for user 3]
%     {\includegraphics[width=0.32\linewidth]{Figures/user3beampattern_cut_1230.eps}\label{fig:u3bp}}
%     \caption{Predictive beam patterns for the three-user case.
% }\label{fig:all_users_bp}
%     \vspace{-1em}
% \end{figure*}

\begin{figure}[t]
    \centering
    \includegraphics[width=0.36\textwidth]{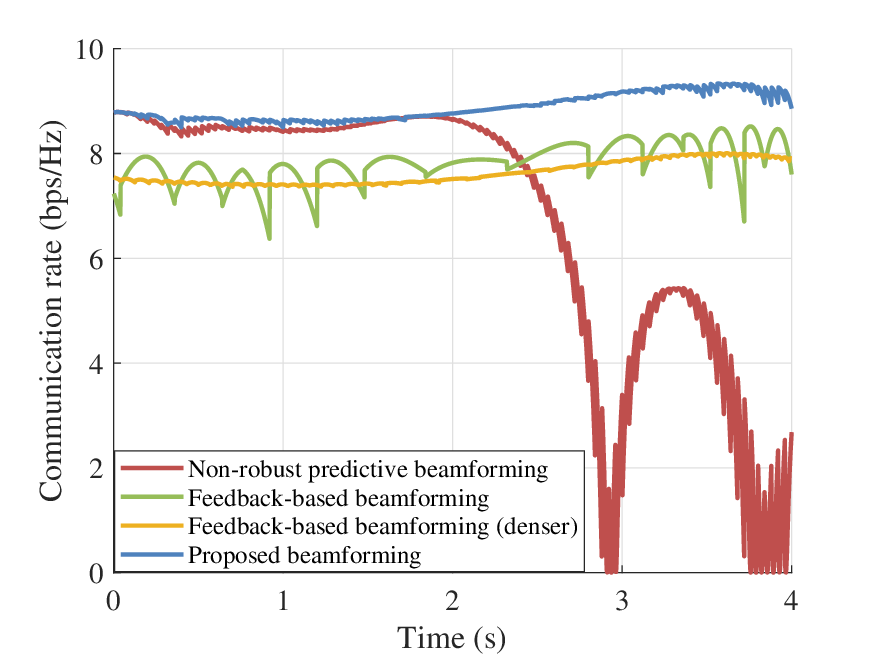}
    \caption{Communication rate over a nonlinear trajectory.}
    \label{fig:rateTime}
    \vspace{-1em}
\end{figure}

\begin{figure}[t]
    \centering
    \includegraphics[width=0.36\textwidth]{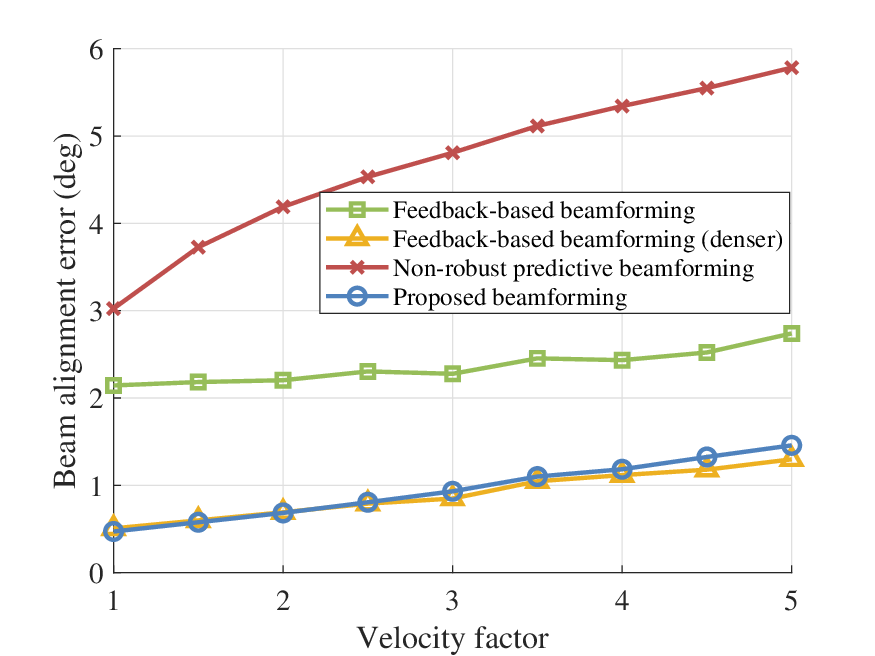}
    \caption{Beam alignment error under different levels of mobility.}
    \label{fig:beamPointingError}
    \vspace{-1em}
\end{figure}

\begin{figure}[t]
    \centering
    \includegraphics[width=0.36\textwidth]{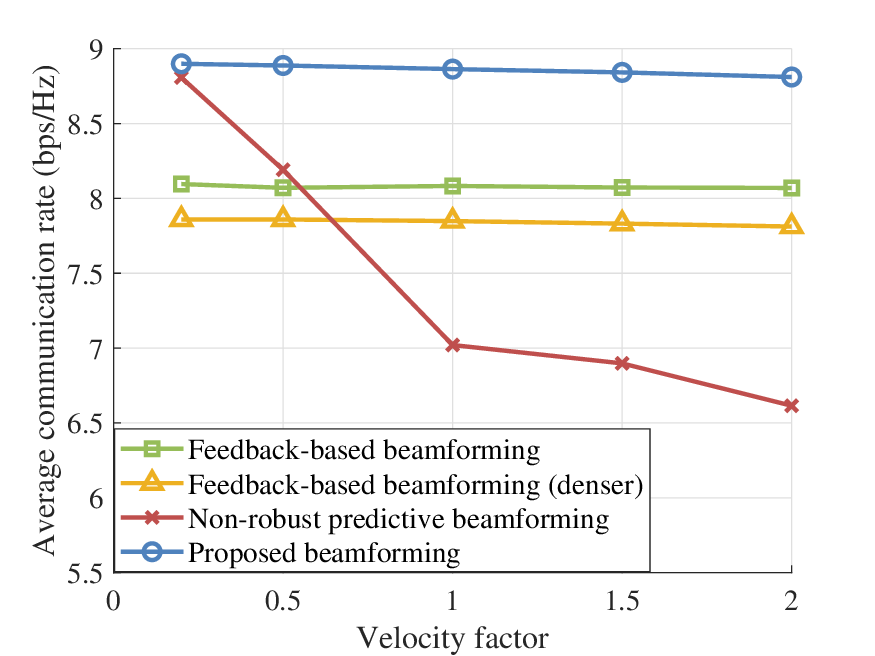}
    \caption{Average communication rate under different levels of mobility.}
    \label{fig:rateVelocity}
    \vspace{-1em}
\end{figure}

The communication rate over this 4-second maneuvering trajectory is evaluated, as illustrated in Fig. \ref{fig:rateTime}. The power budget is set as $P_\mathrm{max} = 10$ dBm. The SSB-based predictive beamforming that does not consider the uncertainty in EKF and beamforming is also adopted as a benchmark (marked as `non-robust predictive beamforming').
%Two types of benchmarks are adopted. For the predicted beamforming method, the BS performs instantaneous beamforming based only on sensing and tracking results without considering the potential uncertainties. 
The proposed predictive robust beamforming method generally yields better performance than feedback-based methods, eliminating the fixed overhead of frequent beam refinement and feedback. 
%It can be inferred from Fig. \ref{fig:trajectory} that the tracking results deviate from the true trajectory after the UAV maneuver, and this affects the subsequent tracking even when the maneuver stops. 
%This phenomenon can be observed in the communication rate performance, where, while the overall rate decreases due to the increased distance between the BS and UAV, a sudden rate drop happens after 3 s and lasts for the entire observation period. 
%As the number of antennas increases, a higher communication rate can be achieved, but the inherent beamwidth of the UPA becomes narrower. Consequently, for the $32 \times 8$ and $32 \times 16$ arrays, 
% The range between the BS and the UAV first decreases and then increases during the 4 seconds, leading to the general {\color{blue} concave} curves. %For the proposed method and predictive method, rate drops happen after acceleration, indicating the model mismatch caused by the EKF. In this case, the proposed method yields better rate performance during maneuvering by incorporating the uncertainties in tracking. 
Since the beamformer produces narrow beams due to the large antenna scale, even a minor feedback lag can cause serious misalignment, leading to rate drops for `feedback-based beamforming', where the refinement is rough and infrequent. However, directly increasing the number of refinements can cause extra overhead, as can be seen from the communication rate given by the `feedback-based beamforming (denser)'.  Although the rate in the `denser' case is more robust with less fluctuation, the achievable rate is, in general, smaller, due to the larger overhead.
In contrast, the non-robust predictive beamforming scheme faces a rate drop after the acceleration, indicating the model mismatch in the EKF.
% In contrast, the proposed beamforming scheme faces a rate drop after the acceleration, indicating the model mismatch in the EKF. 
However, in our proposed beamforming, by incorporating the uncertainties in tracking, the BS provides a relatively wide beam that aims to cover the direction given by the actual trajectory. This gives a robust communication rate.

The impact of UAV's high mobility on beam alignment and communication rate is further evaluated. The velocity factor is defined by temporally scaling the fixed UAV trajectory, such that a larger velocity factor corresponds to traversing the same spatial trajectory in a shorter time. The average beam alignment error under different velocity factors is shown in Fig.~\ref{fig:beamPointingError}. The results show that the beam alignment error generally increases with the velocity factor for all schemes. In particular, the increasing prediction error of the non-robust scheme demonstrates that relying solely on the nominal predicted state becomes less effective as the UAV mobility increases. 
In contrast, the proposed method maintains a relatively low beam alignment error by explicitly accounting for the prediction uncertainty, which further demonstrates the importance of uncertainty-aware beamforming for high-mobility UAV communications. By incorporating the EKF-derived error covariance into the beamforming optimization, the system dynamically achieves coverage robustness, ensuring high reliability in highly mobile aerial environments. The corresponding average communication rate is also demonstrated in Fig.~\ref{fig:rateVelocity} for all cases. Although denser feedback can achieve more accurate beam alignment, the high refinement overhead can lead to a smaller communication rate. As the velocity factor increases, the performance of the non-robust predictive beamforming scheme drops dramatically, while the performance of the other three schemes only decreases slightly.  

\begin{figure}[t]
    \centering
    \includegraphics[width=0.36\textwidth]{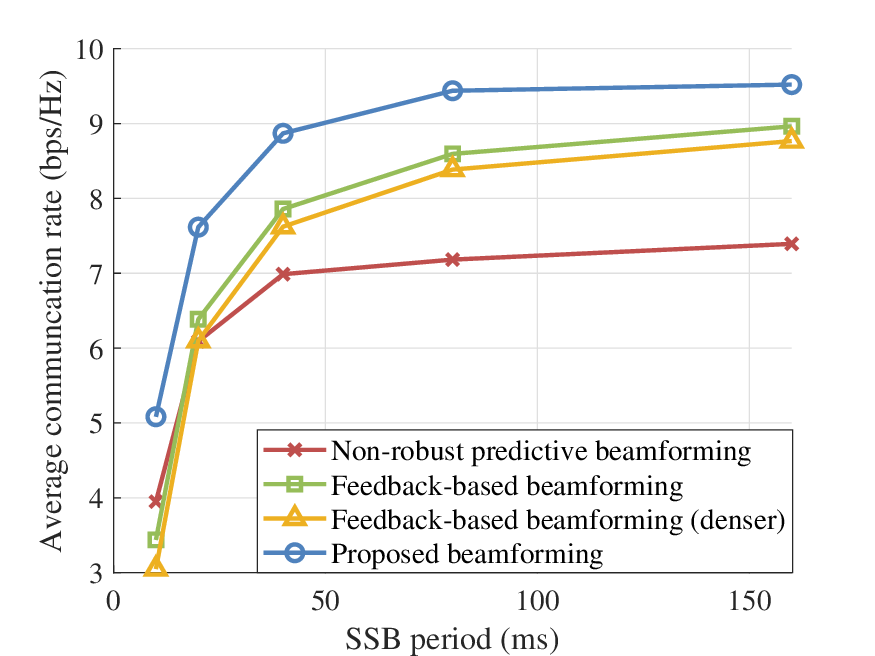}
    \caption{Average communication rate under different SSB periods.}
    \label{fig:ratePeriod}
    \vspace{-1em}
\end{figure}

\begin{figure*}[t]
    \centering
    \subfigure[Beam pattern for user 1]
    {\includegraphics[width = 0.3\linewidth]{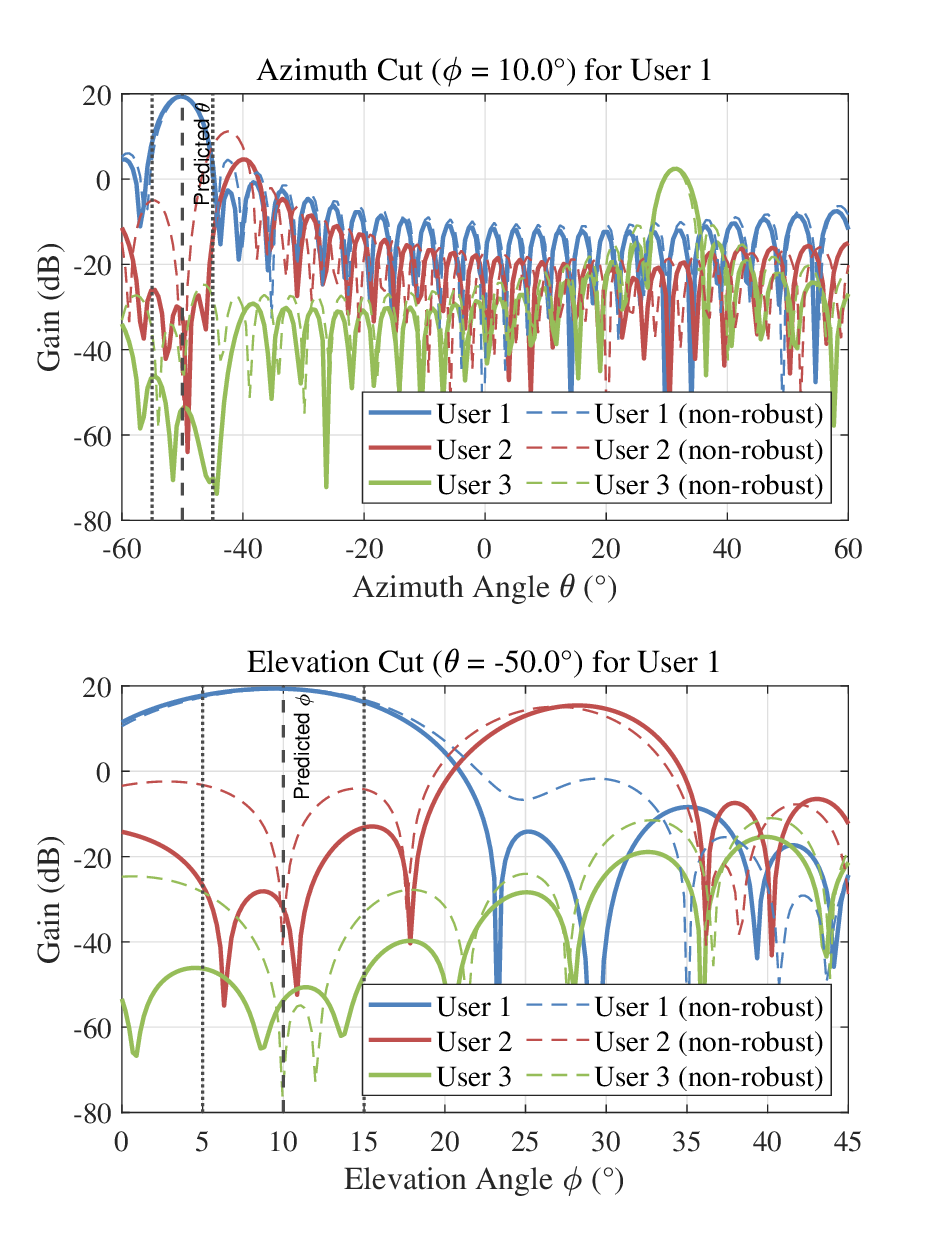}\label{fig:u1bp}}
    \subfigure[Beam pattern for user 2]
    {\includegraphics[width=0.3\textwidth]{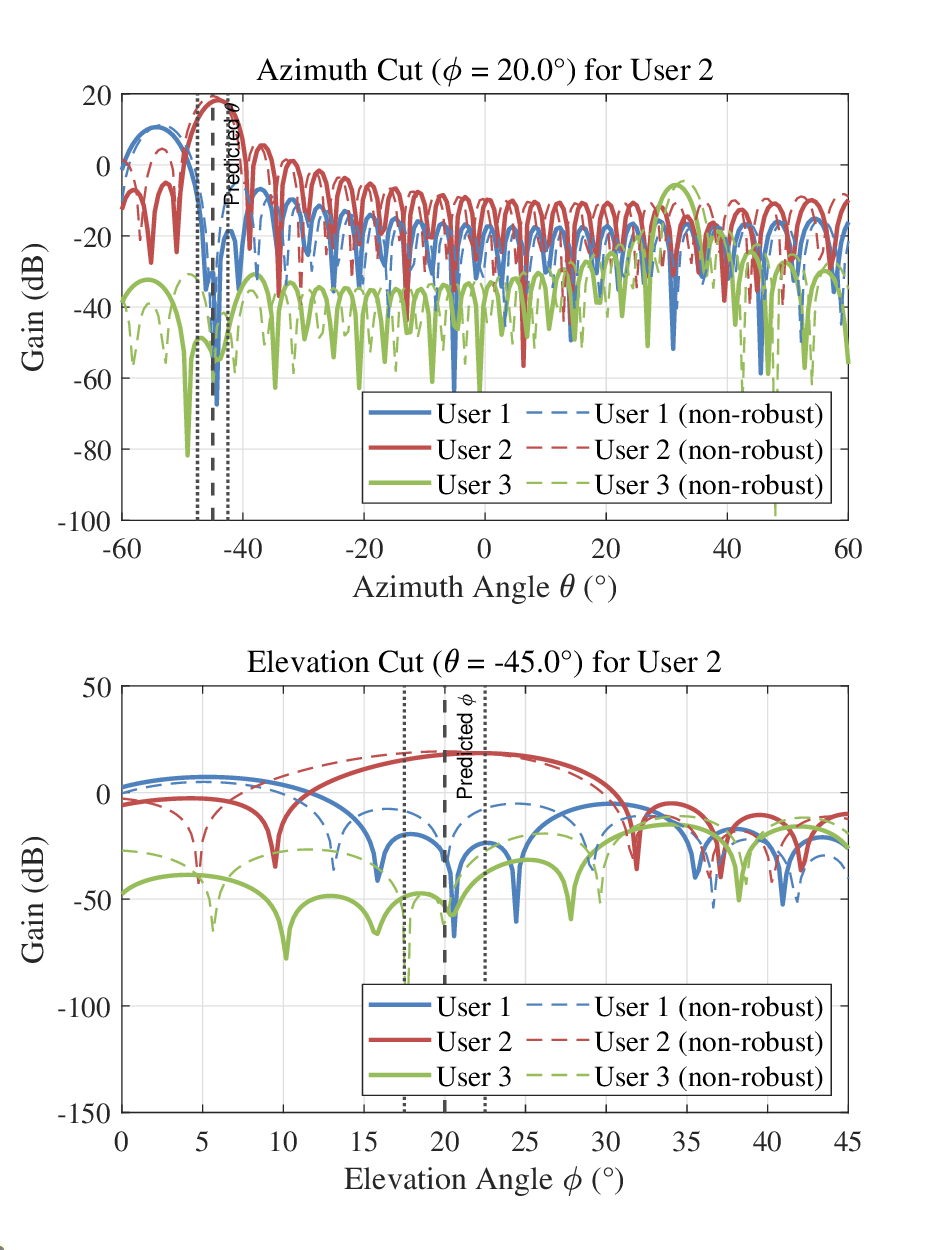}\label{fig:u2bp}}
    \subfigure[Beam pattern for user 3]
    {\includegraphics[width=0.3\linewidth]{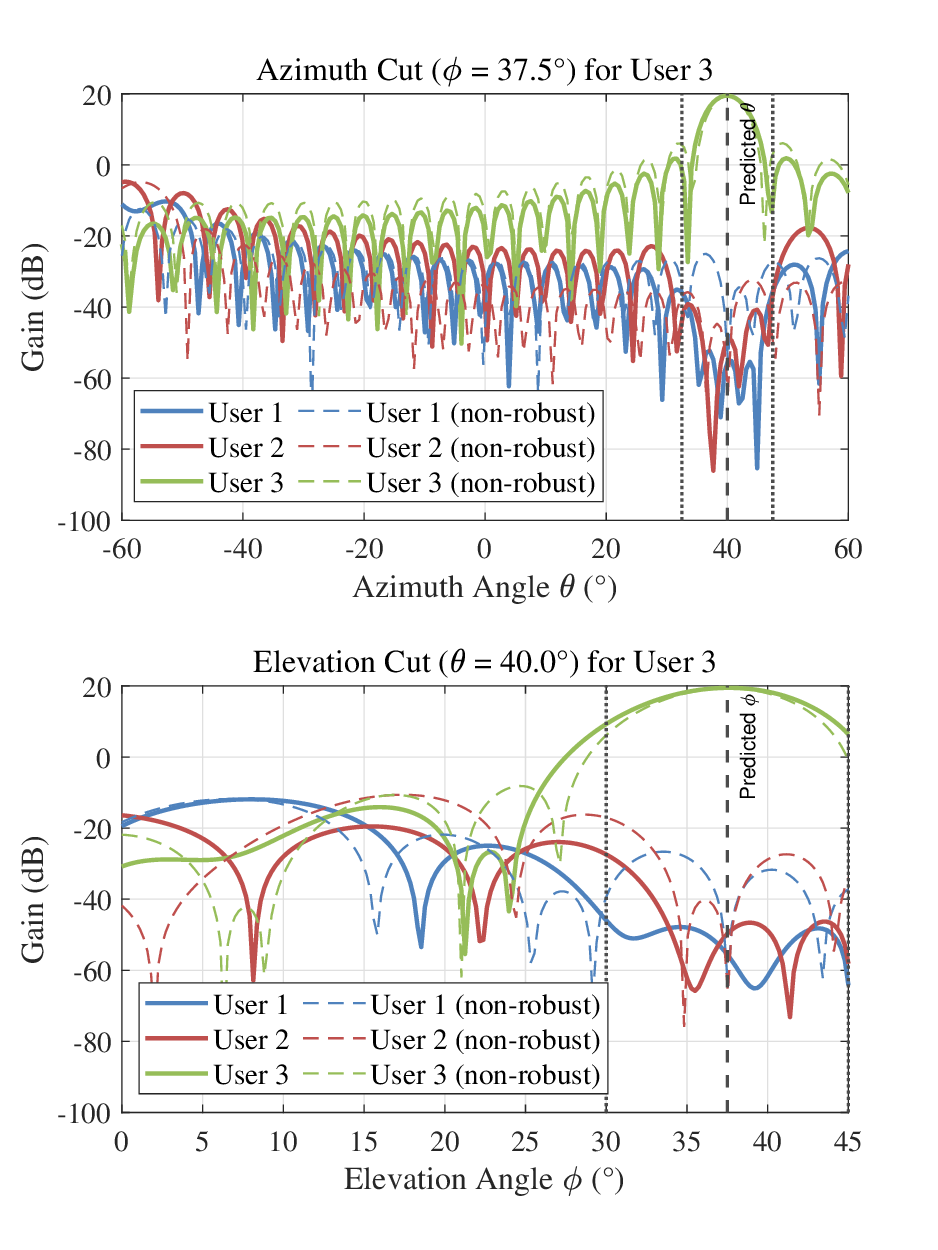}\label{fig:u3bp}}
    \caption{Predictive beam patterns for the three-user case.
}\label{fig:all_users_bp}
    \vspace{-1em}
\end{figure*}

The average communication rate across different SSB measurement periods $T$ ranging from $10$ ms to $160$ ms is further evaluated for a single-UAV case, with a velocity factor equal to 1. As presented in Fig. \ref{fig:ratePeriod}, as the SSB period increases, the average rate for all schemes increases because the SSB training period takes up a smaller proportion of the entire beamforming process. Compared to the feedback-based scheme, the performance gain of our scheme  becomes smaller as the SSB period becomes larger. Although the overhead of SSB is reduced from 80 ms to 160 ms, the communication rates of predictive schemes almost remain the same, due to the increased beam alignment error for a larger SSB period. The EKF tracking performance degrades due to the insufficient number of observations, while the feedback-based methods perform unchanged periodic refinement. As a result, the performance of the non-robust beamforming scheme even becomes worse than the feedback-based schemes for large SSB periods.
We then validate the effectiveness of the proposed beamforming scheme for a multi-user scenario. 
A 3-UAV scenario is considered, with ranges of $[100,200,120]$ m, azimuth angles of $[-50^\circ, -45^\circ, 40^\circ]$, and elevation angles of $[10^\circ,20^\circ,37.5^\circ]$. 
Assume that the uncertainty of the nominal state is characterized by the standard deviations $\sigma$ derived from the corrected state covariance matrix $\tilde{\mathbf{P}}$, which is given by $\sigma_\mathrm{r}=[3.3, 1.7, 0.7] \text{ m}$, $\sigma_\mathrm{\phi}=[1.7^\circ, 0.8^\circ, 2.5^\circ]$, and $\sigma_\mathrm{\theta}=[1.7^\circ, 0.8^\circ, 2.5^\circ]$. These standard deviations possess typical values that are commonly observed in the EKF of different high-mobility scenarios. 
The total transmit power budget is set to $P_{\mathrm{max}} = 30$ dBm, and the average SINR threshold is assumed to be $\gamma_1 = \gamma_2 = \gamma_3 = 15$ dB.

\begin{figure}[t]
    \centering
    \includegraphics[width=0.36\textwidth]{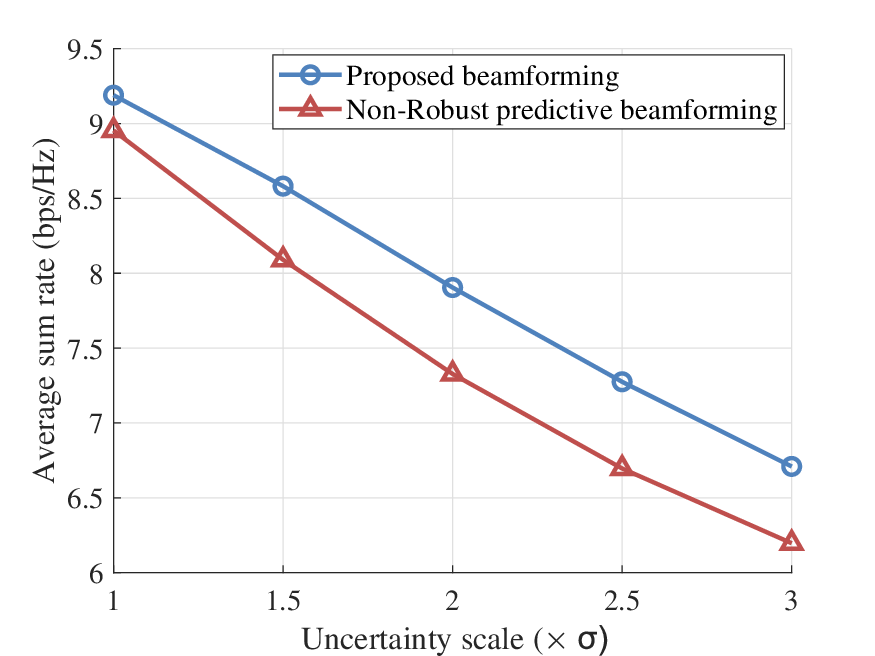}
    \caption{Average sum-rate versus uncertainty scale.}
    \label{fig:sumrate_uncertainty}
    \vspace{-1em}
\end{figure}

The resulting cross-sections of the beam patterns for each user in both azimuth and elevation dimensions are shown in Fig. \ref{fig:all_users_bp}. 
The beam patterns of the non-robust design are also presented, where the non-robust design directly applies the predicted range and angles of each user from the EKF in the beamforming optimization. 
In Fig. \ref{fig:all_users_bp}, the vertical black dashed lines indicate the predicted angular positions obtained from the EKF, while the range between the shaded boundaries represents the $3\sigma$ confidence intervals derived from the estimation uncertainty, covering approximately $99.7\%$ of the target distribution under the Gaussian assumption. As we can see, for the non-robust beamforming design, for a given user, the sidelobes from other users are only well suppressed at the predicted angle. However, our proposed design can achieve a much better sidelobe suppression in the whole uncertainty confidence interval. These results demonstrate that the proposed  beamforming method effectively concentrates the mainlobe gain within the uncertainty region, while the MUI from all other users can be robustly suppressed to improve the SINR, so as to boost the network sum-rate. As a result, by considering the estimation and prediction error of each user/target, our proposed design can achieve a better and more robust performance than the non-robust beamforming design.

To provide a more comprehensive evaluation of the proposed multi-user robust beamforming framework, we further investigate its performance in terms of sum-rate and link reliability. The simulation follows the same 3-UAV setup, with uncertainty levels characterized by the standard deviations $\sigma_\mathrm{r}$, $\sigma_\mathrm{\phi}$, and $\sigma_\mathrm{\theta}$ derived from the corrected EKF covariance matrix. 

\begin{figure}[t]
    \centering
    \includegraphics[width=0.36\textwidth]{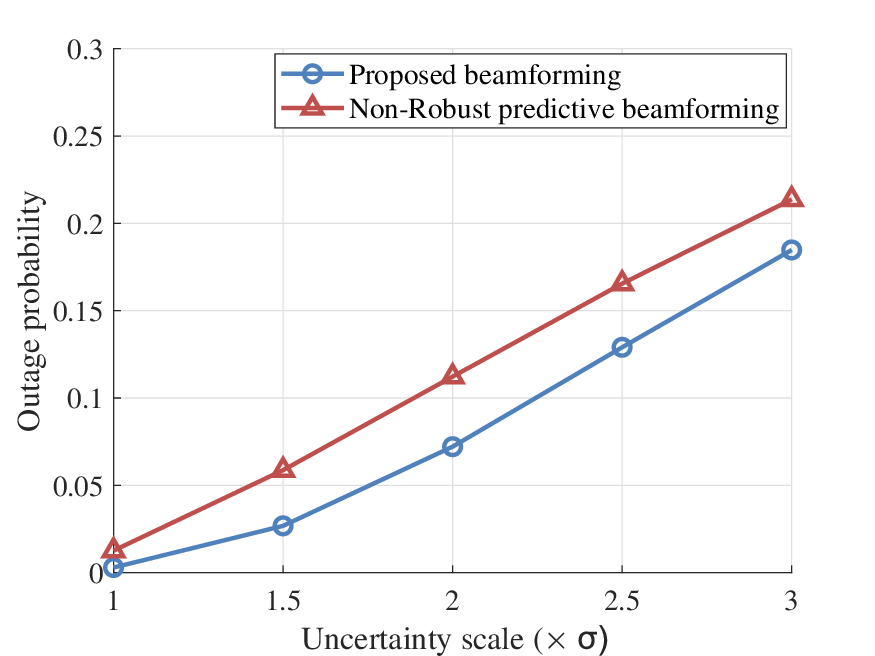}
    \caption{Outage probability versus uncertainty scale.}
    \label{fig:outage_uncertainty}
    \vspace{-1em}
\end{figure}

The average sum-rate with respect to the uncertainty scale is shown in Fig.~\ref{fig:sumrate_uncertainty}. The proposed robust beamforming scheme consistently achieves a higher sum-rate compared with the non-robust design. As the uncertainty scale increases, the performance gap between the two schemes widens, demonstrating that the proposed uncertainty-aware optimization effectively mitigates performance degradation caused by prediction errors. This verifies that incorporating the statistical uncertainty information into the beamforming optimization significantly improves spectral efficiency under imperfect state prediction, which is particularly critical in high-mobility scenarios.

\begin{figure}[t]
    \centering
    \includegraphics[width=0.36\textwidth]{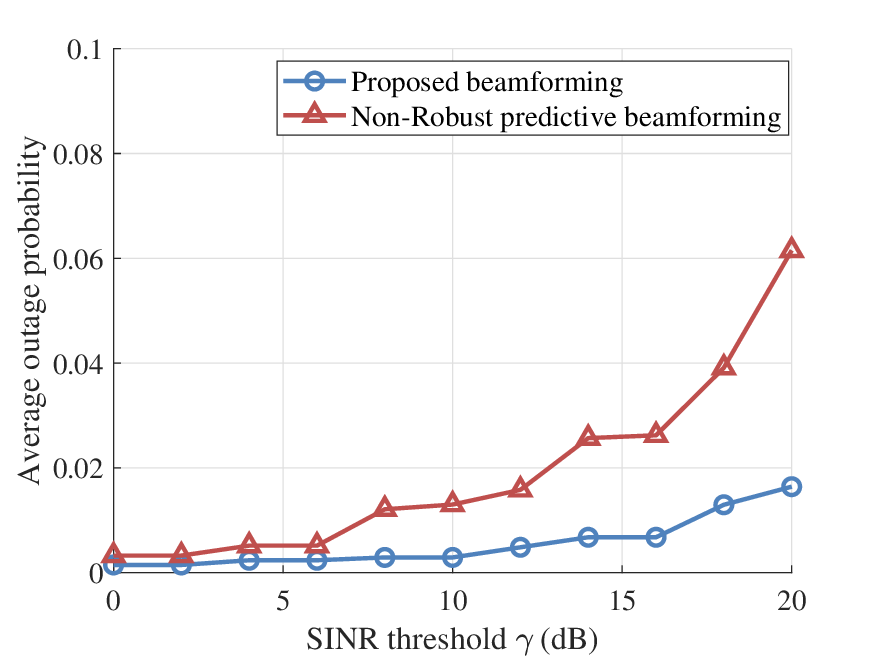}
    \caption{Average outage probability versus SINR threshold.}
    \label{fig:outage_threshold}
    \vspace{-1em}
\end{figure}

The robustness of our design is further evaluated by analyzing the link outage probability, defined as the probability that a user's SINR falls below the required threshold $\gamma$. Fig.~\ref{fig:outage_uncertainty} shows the average outage probability of three UAVs versus the uncertainty scale, with the SINR threshold fixed at $\gamma= 10$ dB. The proposed robust beamforming achieves significantly lower outage probability than the non-robust scheme across all uncertainty levels. Fig.~\ref{fig:outage_threshold} further presents the average outage probability under different SINR thresholds, with uncertainty scale equal to 1. The proposed design maintains a lower outage probability for all considered SINR requirements, confirming its enhanced link reliability in high-mobility multi-user environments. These results highlight that the uncertainty-aware robust beamforming design can effectively guarantee the QoS of individual users even when the state predictions are subject to considerable errors.

\begin{figure}[t]
    \centering
    \subfigure[Average-rate per user ]
    {
        \includegraphics[width=0.45\columnwidth]{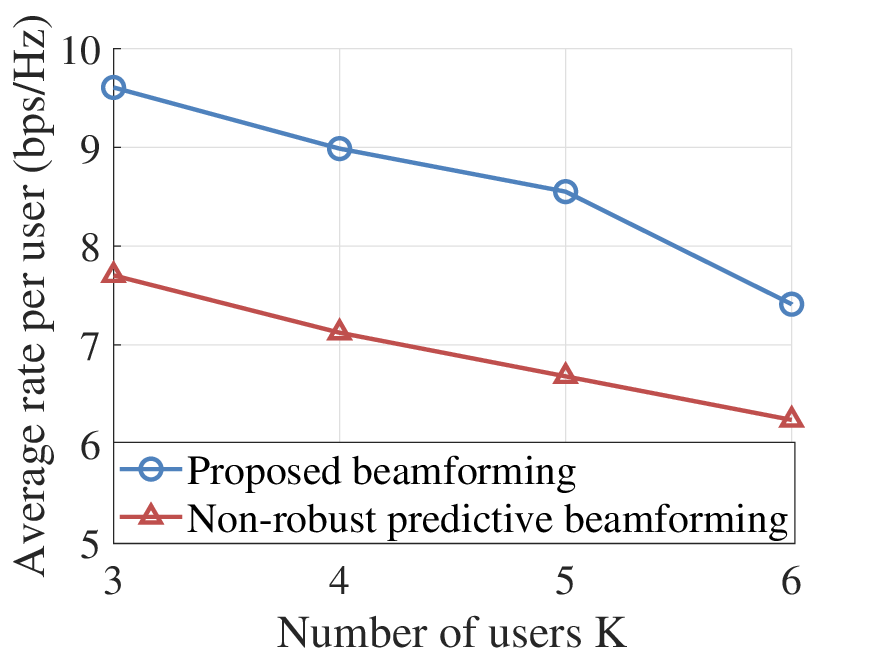}
        \label{fig:rateUserNumber}
    }
    \hfill
    \subfigure[Sum-rate]
    {
        \includegraphics[width=0.45\columnwidth]{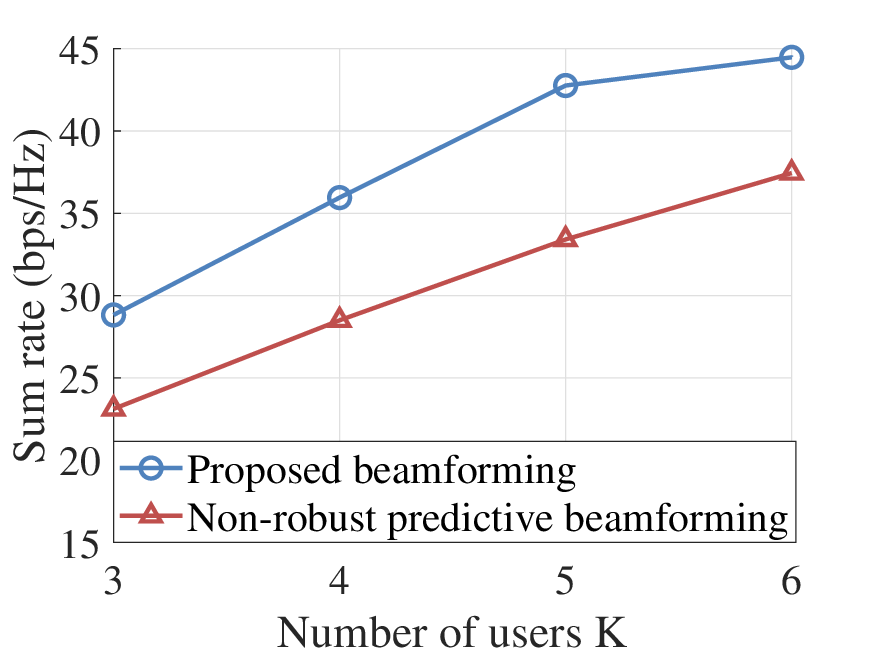}
        \label{fig:sumRateUserNumber}
    }
    \caption{Average and total communication rate versus the number of users.}
    \label{fig:rateComparison}
    \vspace{-1em}
\end{figure}

The rate performance with respect to user number is further evaluated. Specifically, $K=3-6$ UAVs are considered using a predefined set of UAV locations and angular directions. The ranges are selected from [80, 220] m, while the azimuth and elevation angles are configured within the considered coverage region of [-60$^\circ$, 60$^\circ$] and [0$^\circ$, 45$^\circ$], respectively. The actual UAV states used for performance evaluation are perturbed from their predicted states according to the prescribed state errors, thereby accounting for the mismatch between the predicted and actual UAV locations. 
The average per-user communication rate is shown in Fig.~\ref{fig:rateUserNumber}. As $K$ increases, the per-user rate gradually decreases due to the sharing of spatial and power resources and the increased multi-user interference. %Nevertheless, the proposed method maintains reliable per-user performance with up to six simultaneously served UAVs. 
The corresponding total sum-rate is presented in Fig.~\ref{fig:sumRateUserNumber}, which increases with the number of UAVs as the additional users contribute additional achievable rates. Compared to the non-robust design, our proposed design can achieve a performance gain of 19\% $\sim$28\%. 
These results demonstrate that the proposed robust beamforming framework can effectively support an increasing number of simultaneously served UAVs while accounting for their individual channel uncertainties.

\section{Conclusion} \label{sec:conclusion}
%In this paper, a two-phased sensing-assisted multi-user robust beamforming scheme was developed for high-mobility UAV communications in ISAC systems. The sensing phase utilized the SSB to obtain UAV states. In the communication phase, the EKF was employed to predict UAV states, and the critical uncertainty was derived. This information was further integrated into the robust beamforming optimization. Simulation results confirmed the scheme's superiority in reducing overhead and maintaining robust performance for high-mobility multi-user UAV communications. This paper demonstrates that although the SSB has a large repetition interval, it still has great potential to support sensing-assisted predictive beamforming for multi-user fast-moving UAVs with a proper beamforming design. 
%The EKF-driven robust scheme maintains a flatter sum rate curve across the prediction error boundary, effectively mitigating the severe performance degradation experienced by non-robust methods due to CSI prediction errors in high-mobility scenarios.

This paper developed an SSB-based sensing-assisted robust beamforming framework for high-mobility UAV communications in LAWN. By leveraging the periodic SSB as an intrinsic sensing resource, the proposed design eliminates explicit CSI feedback and establishes a predictive beamforming mechanism. A hierarchical sensing algorithm tailored for hybrid digital-analog arrays was first proposed. An EKF was further employed to bridge sparse SSB measurements and enable continuous state prediction. To address maneuver-induced model mismatch, a covariance correction was introduced to characterize statistical uncertainties. With derived uncertainty distributions in predicted range and angular parameters, a multi-user robust beamforming optimization was formulated to maximize the average sum-rate. The nonconvex optimization problem was efficiently solved using SCA and alternating minimization. Simulation results demonstrated that the proposed sensing-assisted design significantly improves spectral efficiency and link stability compared with conventional feedback-based beamforming and non-robust predictive design, particularly in high-mobility and large-SSB-interval scenarios. The proposed framework provides a scalable and practically implementable solution for feedback-free beam management in future UAV-enabled ISAC systems. 

The LoS-dominant channel model adopted in this work is primarily applicable to suburban or relatively open LAWN environments. In dense urban environments with high-rise buildings and strong reflections, NLoS and multipath effects may become significant, and extending the proposed framework to such complex propagation environments is an important direction for future research.

\bibliographystyle{IEEEtran}
\bibliography{reference}

%\begin{comment}

\begin{IEEEbiography}[{\includegraphics[width=1in,height=1.25in,clip,keepaspectratio]{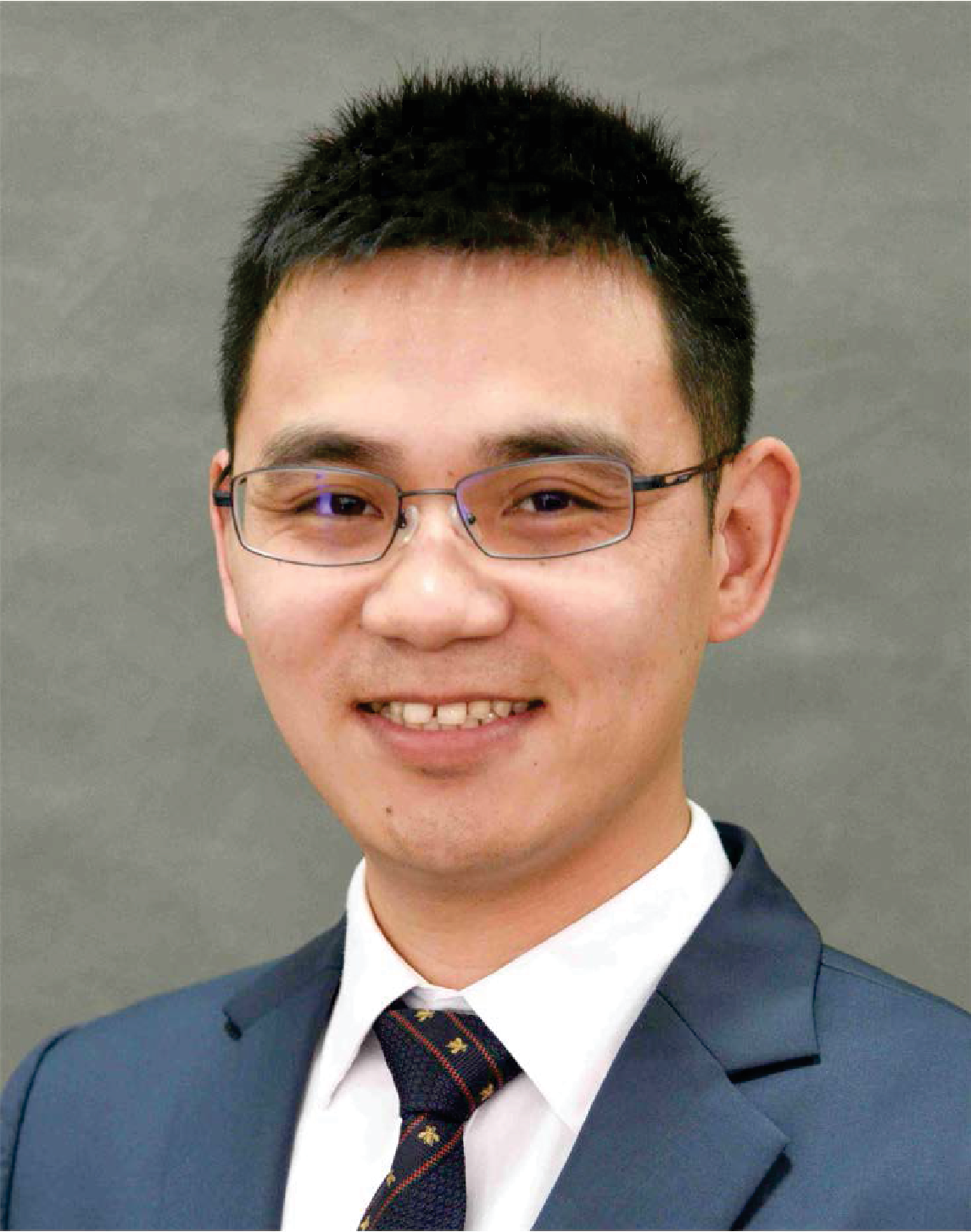}}]{Aimin Tang} (Senior Member, IEEE) received the B.S. and Ph.D. degrees in information and communication engineering from Shanghai Jiao Tong University (SJTU), Shanghai, China, in 2013 and 2018, respectively. He was a visiting Ph.D. student in 2016 at the University of Washington, Seattle, United States. He is an Assistant Research Professor with Shanghai Jiao Tong University. His research interests include B5G/6G networks, integrated sensing and communications, full-duplex communications, and mmWave/THz communications. He received the Outstanding Ph.D. Dissertation Award of SJTU in 2018, the Best Paper Award of IEEE PIMRC in 2021, and the Honorable Mention of MARC Excellent Research Award from MediaTek in 2024. He is currently an Editor of ICT Express. %and had served as the workshop co-chair or technical program co-chair for a few international workshops.
\end{IEEEbiography}

\begin{IEEEbiography}[{\includegraphics[width=1in,height=1.25in,clip,keepaspectratio]{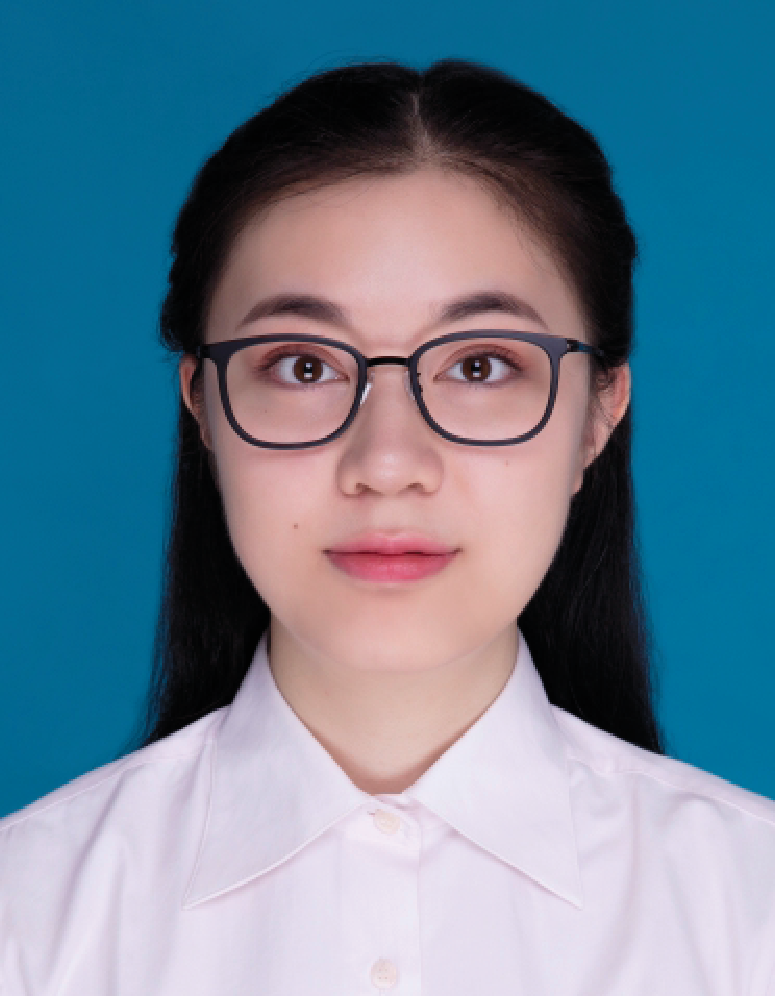}}]{Shuhan Wang} (Graduate Student Member, IEEE) received the B.S. degree in electrical and computer engineering in 2023, and the master's degree in information and communication engineering in 2026, both from Shanghai Jiao Tong University, Shanghai, China. Her research interests include integrated sensing and communications (ISAC) and sensing-assisted communications.
\end{IEEEbiography}

\begin{IEEEbiography}[{\includegraphics[width=1in,height=1.25in,clip,keepaspectratio]{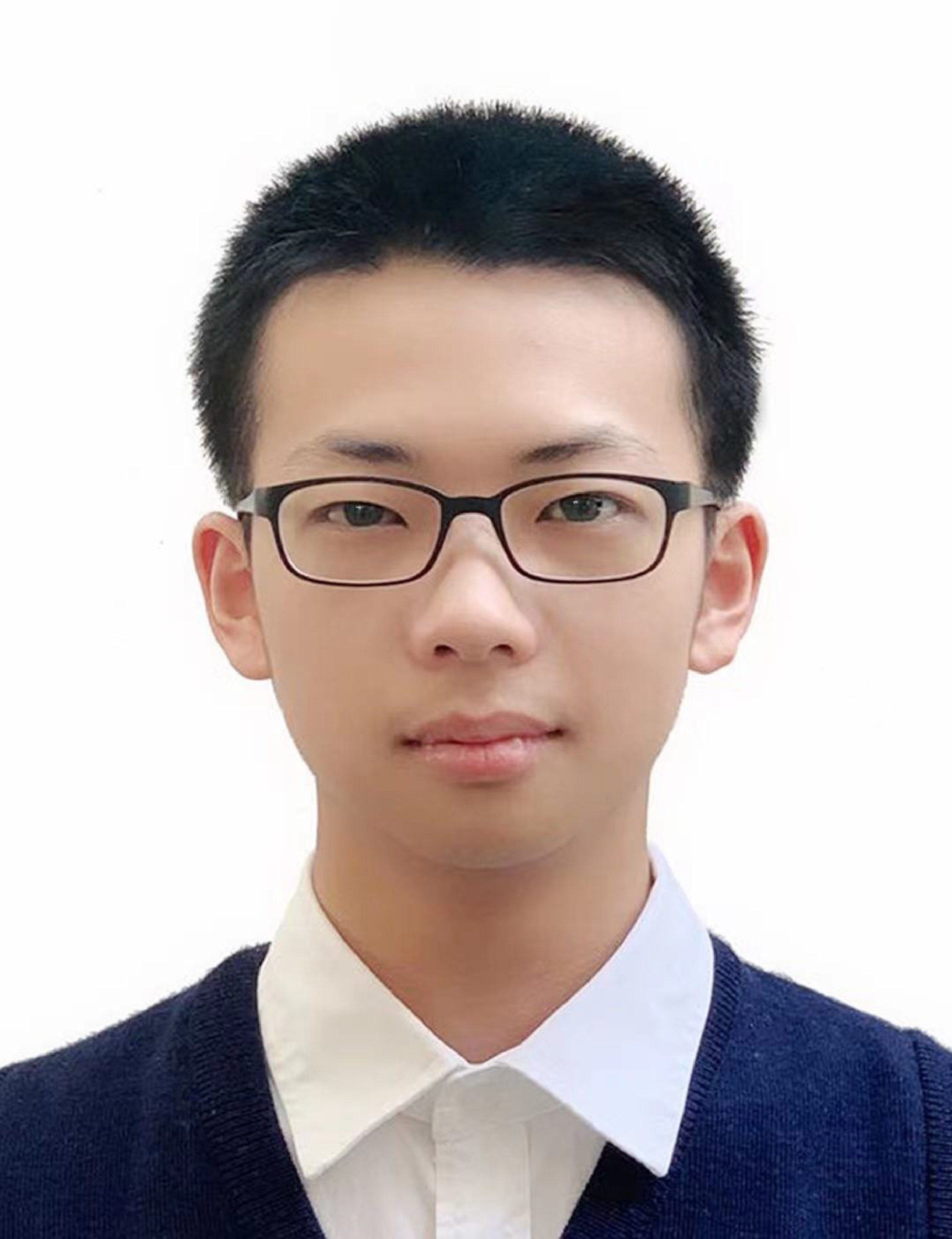}}]{Chongrui Wang}
(Graduate Student Member, IEEE) received the B.S. degree in electrical and computer engineering in 2025 and is currently pursuing the M.S. degree in information and communication engineering, both from Shanghai Jiao Tong University, Shanghai, China. His research interests include signal processing and experimental validation for integrated sensing and communications (ISAC).
\end{IEEEbiography}

\begin{IEEEbiography}[{\includegraphics[width=1in,height=1.25in,clip,keepaspectratio]{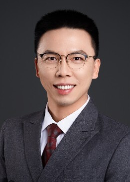}}]{Yin Xu}
(Senior Member, IEEE) is a Full Professor with Shanghai Jiao Tong University. His research interests span key technologies such as channel coding and modulation, new waveforms, inter-tower communications, fluid/movable antenna, and AI-assisted communications, along with the development of prototype systems for diverse communication systems, including 5G/6G, satellite, broadcast, and short-range communications. He was a recipient of the 6G Star Youth Scientist Award. He serves as a co-chair, session chair, and keynote speaker for various major IEEE international conferences. %He is actively involved in communication system standardization, serving as the Chief 3GPP Standard Delegate, representing SJTU, and holding the vice-chair position of Implementation Team 5 under the American Advanced Television Systems Committee (ATSC). His contributions to core physical layer technologies have been incorporated as key components into multiple globally reputable communication standards, including 3GPP 5G, ATSC 3.0, and SparkLink. 
He serves as an Associate Editor for IEEE TRANSACTIONS ON BROADCASTING.
\end{IEEEbiography}
%\end{comment}

\end{document}